\documentclass[11pt]{article}

\usepackage{times}
\usepackage[paperwidth=199.8mm,paperheight=297mm,centering,
            hmargin=20mm,vmargin=20mm]{geometry}
\usepackage[T1]{fontenc}
\usepackage[utf8]{inputenc}
\usepackage{amsmath,amssymb,amsthm,mathtools}
\usepackage{aliascnt}
\usepackage{authblk,microtype,enumitem,needspace}
\usepackage{xcolor}
\usepackage[most]{tcolorbox}
\usepackage{tikz}
\usetikzlibrary{positioning,arrows.meta,calc}
\usepackage{float}
\usepackage[font=small,labelfont=bf,labelsep=period]{caption}
\usepackage{titlesec}

\allowdisplaybreaks[1]

\titleformat{\section}{\normalfont\Large\bfseries}
  {\thesection}{0.7em}{}
\titleformat{\subsection}{\normalfont\large\bfseries}
  {\thesubsection}{0.7em}{}
\titlespacing*{\section}{0pt}{4.5ex plus 1ex minus .3ex}
  {2ex plus .3ex}
\titlespacing*{\subsection}{0pt}{3.25ex plus .8ex minus .2ex}
  {1.4ex plus .2ex}
\setlist[itemize]{topsep=4pt,itemsep=2pt,parsep=0pt}
\setlist[enumerate]{topsep=4pt,itemsep=2pt,parsep=0pt}

\definecolor{paperpurple}{HTML}{583978}
\definecolor{paperpurpleframe}{HTML}{B39EC8}
\definecolor{paperpurplefill}{HTML}{F7F4FA}
\definecolor{papertheoremfill}{HTML}{F5EFFB}
\definecolor{paperblue}{HTML}{244F76}
\definecolor{paperblueframe}{HTML}{8DA9C5}
\definecolor{paperbluefill}{HTML}{E4EEF8}
\definecolor{paperbluealt}{HTML}{2D587C}
\definecolor{paperbluealtframe}{HTML}{9EAFBF}
\definecolor{paperbluealtfill}{HTML}{E8F0F6}
\definecolor{paperslate}{HTML}{3F566B}
\definecolor{paperslateframe}{HTML}{A1AFBD}
\definecolor{paperslatefill}{HTML}{E8EFF5}
\definecolor{papergray}{HTML}{525D67}
\definecolor{papergrayframe}{HTML}{A5AEB7}
\definecolor{papergrayfill}{HTML}{E9EDF1}

\colorlet{paperteal}{paperblue}
\colorlet{papertealframe}{paperblueframe}
\colorlet{papertealfill}{paperbluefill}
\colorlet{papersage}{paperbluealt}
\colorlet{paperborder}{papergrayframe}
\definecolor{paperlink}{HTML}{17436B}
\definecolor{papercite}{HTML}{1C4D79}
\definecolor{paperurl}{HTML}{275378}
\definecolor{paperremarkframe}{HTML}{A9AFB6}
\definecolor{paperremarkfill}{HTML}{EEF0F2}

\usepackage[hypertexnames=false,colorlinks=true,
  linkcolor=paperlink,citecolor=papercite,urlcolor=paperurl]{hyperref}
\DeclareUrlCommand\path{\urlstyle{rm}}
\usepackage[capitalise,noabbrev,nameinlink]{cleveref}
\crefname{boxtheorem}{Theorem}{Theorems}
\Crefname{boxtheorem}{Theorem}{Theorems}
\crefname{boxproposition}{Proposition}{Propositions}
\Crefname{boxproposition}{Proposition}{Propositions}
\crefname{boxcorollary}{Corollary}{Corollaries}
\Crefname{boxcorollary}{Corollary}{Corollaries}
\crefname{boxlemma}{Lemma}{Lemmas}
\Crefname{boxlemma}{Lemma}{Lemmas}
\crefname{boxdefinition}{Definition}{Definitions}
\Crefname{boxdefinition}{Definition}{Definitions}
\crefname{boxassumption}{Assumption}{Assumptions}
\Crefname{boxassumption}{Assumption}{Assumptions}
\crefname{boxremark}{Remark}{Remarks}
\Crefname{boxremark}{Remark}{Remarks}

\newtheoremstyle{paperthmstyle}{0pt}{0pt}{\itshape}{}
  {\bfseries\color{black}}{.}{.6em}{}
\newtheoremstyle{paperpropstyle}{0pt}{0pt}{\itshape}{}
  {\bfseries\color{black}}{.}{.6em}{}
\newtheoremstyle{papercorstyle}{0pt}{0pt}{\itshape}{}
  {\bfseries\color{black}}{.}{.6em}{}
\newtheoremstyle{paperlemstyle}{0pt}{0pt}{\itshape}{}
  {\bfseries\color{black}}{.}{.6em}{}
\newtheoremstyle{paperdefstyle}{0pt}{0pt}{\normalfont}{}
  {\bfseries\color{black}}{.}{.6em}{}
\newtheoremstyle{paperremarkstyle}{0pt}{0pt}{\normalfont}{}
  {\bfseries\color{black}}{.}{.6em}{}

\theoremstyle{paperthmstyle}
\newtheorem{boxtheorem}{Theorem}[section]
\theoremstyle{paperpropstyle}
\newaliascnt{boxproposition}{boxtheorem}
\newtheorem{boxproposition}[boxproposition]{Proposition}
\aliascntresetthe{boxproposition}
\theoremstyle{papercorstyle}
\newaliascnt{boxcorollary}{boxtheorem}
\newtheorem{boxcorollary}[boxcorollary]{Corollary}
\aliascntresetthe{boxcorollary}
\theoremstyle{paperlemstyle}
\newaliascnt{boxlemma}{boxtheorem}
\newtheorem{boxlemma}[boxlemma]{Lemma}
\aliascntresetthe{boxlemma}
\theoremstyle{paperdefstyle}
\newaliascnt{boxdefinition}{boxtheorem}
\newtheorem{boxdefinition}[boxdefinition]{Definition}
\aliascntresetthe{boxdefinition}
\newaliascnt{boxassumption}{boxtheorem}
\newtheorem{boxassumption}[boxassumption]{Assumption}
\aliascntresetthe{boxassumption}
\theoremstyle{paperremarkstyle}
\newaliascnt{boxremark}{boxtheorem}
\newtheorem{boxremark}[boxremark]{Remark}
\aliascntresetthe{boxremark}
\theoremstyle{plain}

\tcbset{academicbox/.style={
  enhanced jigsaw,breakable,rounded corners,
  frame hidden,boxsep=0pt,boxrule=2pt,
  left=12pt,right=12pt,top=12pt,bottom=12pt,
  grow sidewards by=12pt,
  before skip=8pt,after skip=8pt,
  colback=white}}
\tcolorboxenvironment{boxtheorem}{academicbox,
  colback=papertheoremfill}
\tcolorboxenvironment{boxproposition}{academicbox,
  colback=paperbluefill}
\tcolorboxenvironment{boxcorollary}{academicbox,
  colback=paperbluealtfill}
\tcolorboxenvironment{boxlemma}{academicbox,
  colback=paperslatefill}
\tcolorboxenvironment{boxdefinition}{academicbox,
  colback=papergrayfill}
\tcolorboxenvironment{boxassumption}{academicbox,
  colback=papergrayfill}
\tcolorboxenvironment{boxremark}{academicbox,
  colback=paperremarkfill}

\newcommand{\N}{\mathcal N}
\newcommand{\M}{\mathcal M}
\DeclareMathOperator{\Tr}{Tr}
\DeclareMathOperator{\id}{id}
\DeclareMathOperator{\supp}{supp}
\DeclareMathOperator{\ran}{ran}
\DeclareMathOperator{\diag}{diag}

\hypersetup{bookmarksdepth=2}
\newif\ifshowdraftnotes
\showdraftnotestrue

\title{\LARGE \textbf{Finite attainment and efficient computation of regularized quantum channel relative entropy}}
\author[1]{Yiwei Xu}
\author[1]{Kun Fang \thanks{kunfang@cuhk.edu.cn}}
\affil[1]{\small School of Data Science, The Chinese University of Hong Kong, Shenzhen,\protect \\ Guangdong, 518172, China}
\date{\today}

\begin{document}
\maketitle
\begin{abstract}
Quantum channel relative entropy can be nonadditive, making its regularization
essential for asymptotic analysis.
We study when the regularized value is attained at finite blocklength
and how efficiently it can be approximated.
For the first question, we establish a sharp dichotomy: the regularized value is either attained already at the single-copy level, implying additivity at every blocklength, or is never attained at any finite blocklength. Notably, a simple two-copy additivity test distinguishes between these alternatives for
qubit channels and for channel pairs in arbitrary dimensions admitting
an interior optimizer.
We further show that both the interiority condition and finite nonattainment are generic, holding for almost every channel pair.
For the second question, we establish
explicit convergence rates for approximating the regularized channel
relative entropy.
For fixed channel dimensions,
$\widetilde O(\varepsilon^{-3/2})$ channel copies suffice to approximate the regularized value to additive accuracy $\varepsilon$.
Alternatively, a semidefinite hierarchy achieves the same accuracy at
level $\widetilde O(\varepsilon^{-2})$, yielding a certified polynomial-time approximation.
Together, these results reveal the intriguing structure and generic behavior of regularized channel relative entropy, sharpening our understanding of both its analytical and computational aspects.
\end{abstract}

\tableofcontents

\section{Introduction}\label{sec:introduction}

Quantum relative entropy is a fundamental measure of state distinguishability
and a unifying tool in quantum information theory~\cite{Vedral02,Wilde17,Hayashi17}.
Basic entropic quantities, including von Neumann entropy, conditional entropy,
and mutual information, can all be expressed in terms of relative
entropy~\cite{NC10}.
It determines the error exponent in quantum
hypothesis testing~\cite{HP91,ON00}, underlies quantum resource measures~\cite{CG19}, characterizes extractable work in
quantum thermodynamics~\cite{VA16}, and enables numerical estimation in quantum key distribution~\cite{CML16,WLC18}.
These roles motivate extending relative entropy from quantum states to
quantum processes.

A state divergence extends to quantum channels by comparing their outputs on
a common input, which may be entangled with a reference system.
Optimizing the divergence between the resulting joint output states over all
such inputs defines the corresponding channel divergence.
Channel divergences provide tools for channel
discrimination~\cite{WBHK20}, communication over quantum
channels~\cite{CMW16,WFD19,WFT19,FF21GRD}, and resource conversion between channel
pairs~\cite{WW19Distinguishability}.
Here we study the Umegaki channel relative entropy,
defined for channels $\N,\M:A\to B$ by
\begin{align}\label{eq:channel-relative-entropy}
D(\N\|\M)
&:=\sup_{\psi_{RA}}D\bigl(
(\id_R\otimes\N)(\psi_{RA})\big\|
(\id_R\otimes\M)(\psi_{RA})\bigr),
\end{align}
where the supremum ranges over joint input states $\psi_{RA}$ and
finite-dimensional reference systems $R$.

A crucial difference from state relative entropy emerges when channels
are used repeatedly.
Whereas state relative entropy is additive on tensor products, entangling
inputs across channel uses can increase the relative entropy per copy.
Channel relative entropy can therefore be strictly
superadditive~\cite{FFRS}, making regularization essential for asymptotic
characterization.
The relevant quantity is the regularized channel relative entropy,
\begin{align}\label{eq:channel-relative-entropy-regularized}
D^{\mathrm{reg}}(\N\|\M)
&:=\sup_{m\ge1}\frac1m
D(\N^{\otimes m}\|\M^{\otimes m}).
\end{align}
This regularized value determines the optimal asymptotic error
exponent in asymmetric channel discrimination~\cite{WW19Distinguishability,FFRS}.
Its operational role motivates two questions about its structure
and computational accessibility:
\begin{tcolorbox}[academicbox,colback=papergrayfill,breakable=false,
  top=8pt,bottom=8pt,before skip=9pt,after skip=9pt]
\begin{enumerate}[label=\textbf{(Q\arabic*)},labelindent=0pt,leftmargin=*,
  labelsep=.6em,itemsep=6pt,topsep=0pt,parsep=0pt]
\item \textbf{Finite attainment.} Can the regularized value be attained
exactly by a finite $m$?
\item \textbf{Efficient approximation.} How can the regularized value
be computed efficiently?
\end{enumerate}
\end{tcolorbox}

Progress on finite attainment has centered on structured channel pairs.
One-copy formulas are known for classical--quantum pairs with a common
classical input basis~\cite{WBHK20} and for discrimination against a replacer
channel~\cite{CMW16}.
They also hold for jointly teleportation-simulable pairs whose resource
states are their Choi states, including suitable jointly covariant
pairs~\cite{WBHK20}.
These results identify settings in which regularization collapses.
The known nonadditive examples~\cite{FFRS}, however, leave open
whether the regularized value might first be attained at a larger
finite blocklength.
They also leave open whether finite nonattainment is exceptional
or typical among quantum channel pairs.

On the computational side, convex optimization provides a convergent
upper hierarchy for regularized sandwiched R\'enyi channel divergences
at orders above one~\cite{FF21}.
Permutation symmetry makes this hierarchy efficient in fixed
dimensions~\cite{FST22}.
Recent work established continuity of these regularized divergences
as the order approaches one~\cite{SS26,WY26,GJL26,ZW26}.
In particular, Schmitt and Sutter used this continuity to prove
computability of the regularized channel relative
entropy~\cite[Corollary~4.1]{SS26}.
A quantitative question remains: how should the R\'enyi order and
hierarchy level depend on the requested accuracy?
Answering this question, and controlling finite-block convergence
directly, is essential for obtaining explicit approximation and
complexity guarantees.

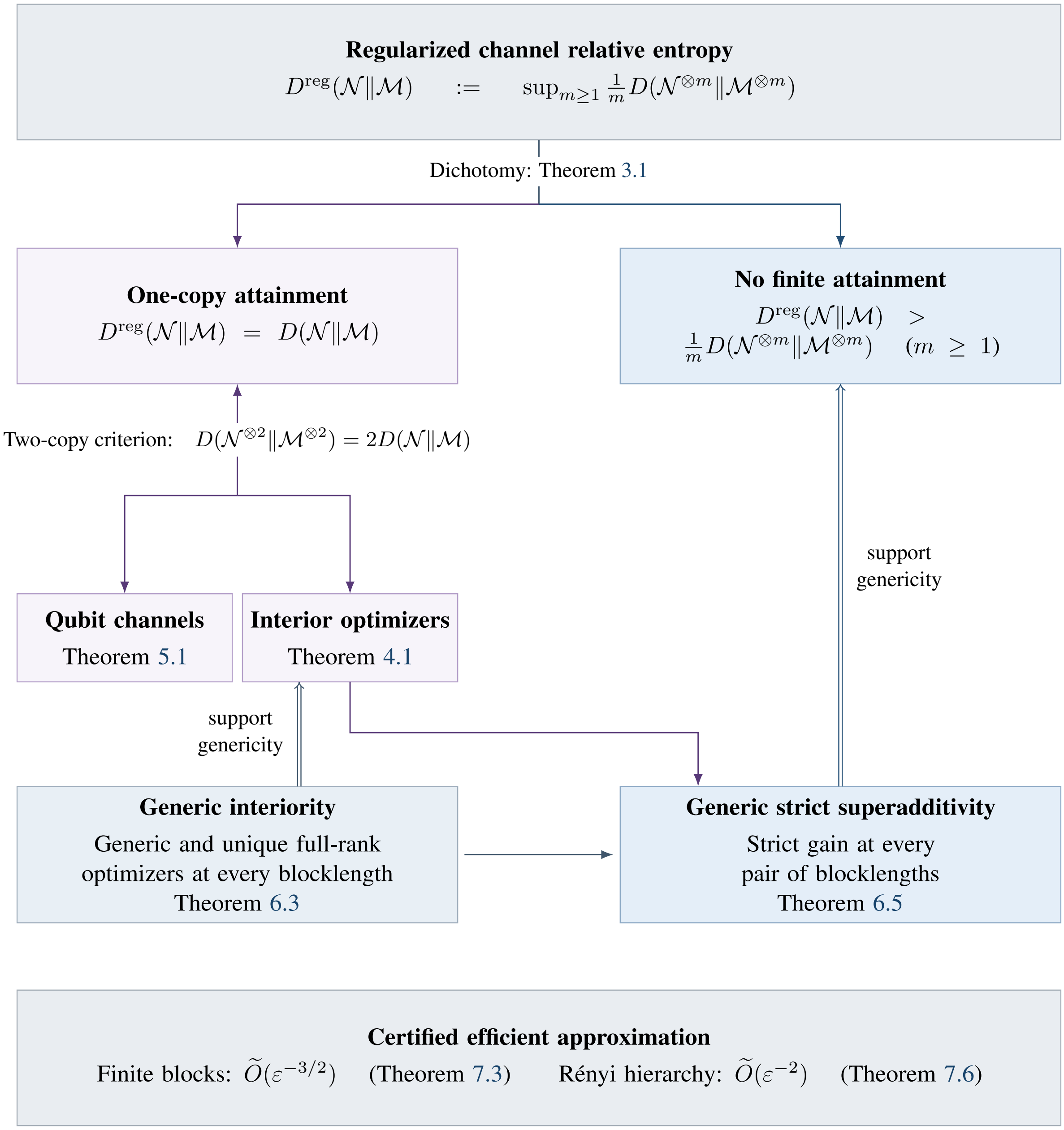
\begin{figure}[!ht]
\centering
\begin{tikzpicture}[
 font=\small,
 box/.style={draw,sharp corners,align=center,
   inner xsep=5pt,inner ysep=5pt,line width=.45pt},
 mainbox/.style={box,text width=6.15cm,minimum height=20mm},
 root/.style={mainbox,draw=papergrayframe,fill=papergrayfill,
   text width=15.05cm},
 attained/.style={mainbox,draw=paperpurpleframe,fill=paperpurplefill},
 unattained/.style={mainbox,draw=paperblueframe,fill=paperbluefill},
 criterion/.style={box,draw=paperpurpleframe,fill=paperpurplefill,
   text width=3cm,inner xsep=2.5pt,minimum height=13mm},
 generic/.style={mainbox,draw=paperslateframe,fill=paperslatefill},
 algorithm/.style={box,draw=papergrayframe,fill=papergrayfill,
   text width=15.05cm,minimum height=20mm},
 arrow/.style={-{Latex[length=1.8mm]},line width=.55pt},
 support/.style={double=white,double distance=1pt,
   -{Implies[length=2.2mm,width=2mm]},line width=.45pt},
 equiv/.style={{Latex[length=1.6mm]}-{Latex[length=1.6mm]},
   line width=.55pt},
 every node/.style={outer sep=0pt}
]

\node[root] (top) at (0,1.15)
 {\textbf{Regularized channel relative entropy}\\[3pt]
  $D^{\mathrm{reg}}(\N\|\M):=\sup_{m\ge1}\frac1m
D(\N^{\otimes m}\|\M^{\otimes m})$};

\node[attained] (yes) at (-4.45,-2.45)
 {\textbf{One-copy attainment}\\[3pt]
  $D^{\mathrm{reg}}(\N\|\M)=D(\N\|\M)$};

\node[unattained] (no) at (4.45,-2.45)
 {\textbf{No finite attainment}\\[3pt]
  $D^{\mathrm{reg}}(\N\|\M)>
\frac1mD(\N^{\otimes m}\|\M^{\otimes m})$
  \quad($m\ge1$)};

\node[criterion,anchor=west] (qubit) at ($(yes.west)+(0,-4.75)$)
 {\textbf{Qubit channels}\\[3pt]
  Theorem~\ref{thm:qubit}};

\node[criterion,anchor=east] (interior) at ($(yes.east)+(0,-4.75)$)
 {\textbf{Interior optimizers}\\[3pt]
  Theorem~\ref{thm:modular}};

\node[generic] (genint) at (-4.45,-10.40)
 {\textbf{Generic interiority}\\[3pt]
  Generic and unique full-rank \\
  optimizers at every blocklength\\
  Theorem~\ref{thm:generic-interior}};

\node[unattained] (genstrict) at (4.45,-10.40)
 {\textbf{Generic strict superadditivity}\\[3pt]
  Strict gain at every\\
  pair of blocklengths\\
  Theorem~\ref{thm:generic-superadditivity}};

\node[algorithm] (comp) at (0,-13.40)
 {\textbf{Certified efficient approximation}\\[3pt]
  Finite blocks: $\widetilde O(\varepsilon^{-3/2})$
  \quad(Theorem~\ref{thm:physical-block})\qquad
  R\'enyi hierarchy: $\widetilde O(\varepsilon^{-2})$
  \quad(Theorem~\ref{thm:upper-hierarchy})};

\coordinate (attainment-split) at ($(top.south)+(0,-.95)$);
\draw[draw=paperslate,line width=.55pt]
  (top.south)
  --
  node[midway,anchor=center,font=\footnotesize,text=black,fill=white,
    inner xsep=3pt,inner ysep=2pt]
    {Dichotomy: Theorem~\ref{thm:universal}}
  (attainment-split);
\draw[arrow,draw=paperpurple]
  (attainment-split) -| (yes.north);
\draw[arrow,draw=paperblue]
  (attainment-split) -| (no.north);

\coordinate (criterion-split) at ($(yes.south)+(0,-1.65)$);
\draw[{Latex[length=1.8mm]}-,draw=paperpurple,line width=.55pt]
  (yes.south)
  --
  node[midway,anchor=center,overlay,font=\footnotesize,text=black,fill=white,
    inner xsep=3pt,inner ysep=2pt]
    {Two-copy criterion:\quad
     $D(\N^{\otimes 2}\|\M^{\otimes 2})=2D(\N\|\M)$}
  (criterion-split);
\draw[arrow,draw=paperpurple]
  (criterion-split) -| (qubit.north);
\draw[arrow,draw=paperpurple]
  (criterion-split) -| (interior.north);

\draw[arrow,draw=paperpurple]
  (interior.south)
  -- ++(0,-.75)
  -|
  ($(genstrict.north west)+(1.15,0)$);

\draw[arrow,draw=paperslate]
  ($(genint.east)+(0.10,0)$)
  --
  ($(genstrict.west)+(-0.10,0)$);

\coordinate (interiority-support) at ($(interior.south)+(-.75,0)$);
\draw[support,draw=paperslate]
  (genint.north -| interiority-support)
  --
  node[midway,left=3pt,font=\footnotesize,align=center]{support\\genericity}
  (interiority-support);

\draw[support,draw=paperblue]
  (genstrict.north)
  --
  node[pos=.54,right=3pt,font=\footnotesize,align=center]{support\\genericity}
  (no.south);

\end{tikzpicture}

\caption{Overview of the results.
Single-line, one-way arrows indicate proof dependencies,
while bidirectional arrows denote equivalences.
Double-line arrows indicate that the corresponding properties support the genericity.
The finite-attainment results assume $D(\N\|\M)<\infty$,
while the approximation bounds require $J_{\M}>0$.
The stated approximation orders refer to blocklengths
and hierarchy levels.}
\label{fig:overview}
\end{figure}

In this work, we address these questions through four main contributions. An overview of the main results and their interconnections is given in
Figure~\ref{fig:overview}. 
\begin{itemize}

\item \textbf{Finite-attainment dichotomy.}
For every channel pair, we find that
the finite attainment of the regularized quantum channel relative entropy
admits a dichotomy (Theorem~\ref{thm:universal}): either the regularized value is
attained at one copy, implying additivity at every
blocklength, or it is never attained at any finite
blocklength. That is, exactly one of the following holds:
\begin{align}
D^{\mathrm{reg}}(\N\|\M)=D(\N\|\M) \quad\text{or}\quad 
D^{\mathrm{reg}}(\N\|\M)>
\frac1mD(\N^{\otimes m}\|\M^{\otimes m}),\ \forall m.
\end{align}

\item \textbf{Two-copy criteria.}
Determining which case of the dichotomy holds by directly comparing
$D^{\mathrm{reg}}(\N\|\M)$ with $D(\N\|\M)$ is difficult because the
regularized divergence involves an asymptotic optimization.
We show that this comparison reduces to a two-copy additivity test:
\begin{align}
D^{\mathrm{reg}}(\N\|\M)=D(\N\|\M) \iff D(\N^{\otimes 2}\|\M^{\otimes 2})=2D(\N\|\M).
\end{align}
The equivalence holds for any pair of qubit channels
(Theorem~\ref{thm:qubit}) and for channel pairs in arbitrary dimensions
admitting an interior optimizer (Theorem~\ref{thm:modular}).
Here, interiority means that some optimal pure one-copy input has a
full-rank reference marginal.

\item \textbf{Generic interiority and finite nonattainment.}
We show that the interiority assumption in the two-copy criterion holds
generically (Theorem~\ref{thm:generic-interior}).
We also establish generic strict superadditivity of channel relative
entropy (Theorem~\ref{thm:generic-superadditivity}),
\begin{align}
  D(\N^{\otimes(m+n)}\|\M^{\otimes(m+n)})
>D(\N^{\otimes m}\|\M^{\otimes m})
+D(\N^{\otimes n}\|\M^{\otimes n}),\qquad m,n\ge1
\end{align} implying that the
regularized value is unattainable at any finite blocklength for almost
every channel pair.
Both genericity statements follow by showing that the exceptional pairs
lie in a proper real algebraic subset of the affine space of
trace-preserving Choi pairs.
Such subsets have Lebesgue measure zero in the ambient affine space.

\item \textbf{Certified efficient approximation.}
When the second channel has a positive-definite Choi matrix, we establish
explicit convergence rates for approximating the regularized channel
relative entropy.
For fixed dimensions and a fixed positive spectral lower bound,
$m = \widetilde O(\varepsilon^{-3/2})$ channel copies\footnote{The notation
$\widetilde O$ suppresses polylogarithmic factors in $1/\varepsilon$.
More precisely, $\widetilde O(\varepsilon^{-p})$ denotes a bound of
the form $O(\varepsilon^{-p}[\log(1/\varepsilon)]^q)$ as
$\varepsilon\to0^+$ for some fixed $q\ge0$.}
suffice to achieve additive accuracy $\varepsilon$
(Theorem~\ref{thm:physical-block}):
\begin{align}
  0&\le D^{\mathrm{reg}}(\N\|\M)
-\frac1mD(\N^{\otimes m}\|\M^{\otimes m})
<\varepsilon.
\end{align}
Alternatively, a semidefinite hierarchy achieves the same accuracy at
level $\widetilde O(\varepsilon^{-2})$, yielding certified polynomial-time
approximation of the regularized value (Theorem~\ref{thm:upper-hierarchy}).

\end{itemize}

Together, these results reveal the intriguing structures and generic behaviors of the regularized quantum channel relative entropy, sharpening our understanding of this fundamental quantity in both analytical and computational aspects.

\paragraph{Organization.} Section~\ref{sec:preliminaries} collects preliminaries, and
Section~\ref{sec:universal} establishes the finite-attainment
dichotomy. Sections~\ref{sec:interior-modular} and~\ref{sec:qubit}
develop the two-copy criteria for interior optimizers and
qubit channels, respectively. Section~\ref{sec:generic}
establishes generic interiority, uniqueness, and strict
superadditivity. Section~\ref{sec:regularized-computation}
presents certified approximation methods and uniform continuity.
Appendix~\ref{app:boundary} provides the boundary perturbation proofs supporting
Sections~\ref{sec:qubit} and~\ref{sec:generic}.

\section{Preliminaries}\label{sec:preliminaries}

\paragraph{Notation.}
For a finite-dimensional Hilbert space $\mathcal H$, let
$\mathcal B(\mathcal H)$ and $\mathcal D(\mathcal H)$ denote
its linear operators and density operators, respectively.
For a system $A$, write $d_A:=|A|$ for its dimension and
$R\simeq A$ when $|R|=|A|$.
We write $X\ge0$ and $X>0$ for positive semidefinite and positive
definite operators, respectively. Operator inequalities are understood
in the L\"owner order.
For $X\ge0$, let $X^+$ denote its Moore--Penrose inverse.
We use $\log$ for base-two logarithms and $\ln$ for natural logarithms.
Relative entropies, R\'enyi divergences, and entropy functions are measured in nats.
We adopt the convention $0\log 0=0\ln 0=0$ and evaluate
matrix logarithms of $X$ on its support, denoted by $\operatorname{supp}X$.

For positive semidefinite operators $X,Y$ with
$\supp X\subseteq\supp Y$, we use the trace-corrected
relative entropy
\begin{align}
D(X\|Y):=\Tr X(\ln X-\ln Y)-\Tr X+\Tr Y,
\end{align}
and set $D(X\|Y)=+\infty$ otherwise.
For states, the trace correction vanishes.

For a channel $\mathcal L:A\to B$ with input dimension $d_A$,
denote its unnormalized Choi matrix by
\begin{align}
J_{\mathcal L}
=(\id\otimes\mathcal L)(|\Phi_{d_A}\rangle\langle\Phi_{d_A}|),
\end{align}
where $|\Phi_{d_A}\rangle=\sum_{j=1}^{d_A}|jj\rangle$ is the unnormalized maximally entangled state.

\paragraph{Regularization and input optimization.}
Product inputs give
\begin{align}
D(\N^{\otimes(m+n)}\|\M^{\otimes(m+n)})
\ge D(\N^{\otimes m}\|\M^{\otimes m})
+D(\N^{\otimes n}\|\M^{\otimes n}).
\end{align}
By Fekete's lemma,
\begin{align}
D^{\mathrm{reg}}(\N\|\M)
=\lim_{m\to\infty}\frac1m
D(\N^{\otimes m}\|\M^{\otimes m}).
\end{align}

In finite dimensions, $D(\N\|\M)<\infty$ is equivalent
to $\supp J_{\N}\subseteq\supp J_{\M}$, and hence to
$J_{\N}\le cJ_{\M}$ for some finite $c$.
Consequently,
\begin{align}\label{eq:choi-domination-bound}
D(\N^{\otimes m}\|\M^{\otimes m})\le m\ln c
\qquad(m\ge1).
\end{align}

By purification and Schmidt decomposition, the optimization
defining $D(\N^{\otimes m}\|\M^{\otimes m})$ reduces to
pure inputs with $R\simeq A^{\otimes m}$.

For a reference marginal $\tau\in\mathcal D(A^{\otimes m})$,
define the $m$-copy objective by
\begin{align}\label{eq:input-objective}
D(\N^{\otimes m},\M^{\otimes m};\tau)
:=D\Bigl(
(\sqrt\tau\otimes I)J_{\N}^{\otimes m}(\sqrt\tau\otimes I)
\Big\|
(\sqrt\tau\otimes I)J_{\M}^{\otimes m}(\sqrt\tau\otimes I)
\Bigr).
\end{align}
For finite $D(\N\|\M)$, this function is concave and
continuous on $\mathcal D(A^{\otimes m})$, and
\begin{align}\label{eq:input-optimization}
D(\N^{\otimes m}\|\M^{\otimes m})
=\max_{\tau\in\mathcal D(A^{\otimes m})}
D(\N^{\otimes m},\M^{\otimes m};\tau).
\end{align}
Here $\tau$ is the reference marginal, while the
channel-input marginal is $\tau^{\mathsf T}$.
Concavity follows from data processing;
see \cite[proof of Proposition~13]{WWS}.

\section{Finite-attainment dichotomy}
\label{sec:universal}

In this section, we show that the finite attainment of the regularized quantum channel relative entropy
admits a dichotomy: either the regularized value is
attained at one copy, implying additivity at every
blocklength, or it is never attained at any finite
blocklength.

\begin{boxtheorem}[Finite-attainment dichotomy]
\label{thm:universal}
Let $A,B$ be finite-dimensional systems, and let
$\N,\M$ be channels from $A$ to $B$ with
$D(\N\|\M)<\infty$.
Then exactly one of the following holds:
\begin{enumerate}[label=\textnormal{(\roman*)}]
\item \emph{One-copy attainment.}
\begin{align}
D^{\mathrm{reg}}(\N\|\M)=D(\N\|\M).
\end{align}
\item \emph{No finite attainment.} For every $m\ge1$,
\begin{align}
D^{\mathrm{reg}}(\N\|\M)>
\frac1mD(\N^{\otimes m}\|\M^{\otimes m}).
\end{align}
\end{enumerate}
\end{boxtheorem}

\begin{proof}
Dividing \eqref{eq:choi-domination-bound} by the blocklength and
taking the supremum gives $D^{\mathrm{reg}}(\N\|\M)\le\ln c<\infty$.
Fix an integer $n\ge2$ and a state $\psi_{RA_1\cdots A_n}$,
where $R$ is a finite reference system and each $A_i\simeq A$. Set
\begin{align}
\rho_k=(\id_R\otimes\N^{\otimes k}\otimes\id)(\psi),
\qquad
\sigma_k=(\id_R\otimes\M^{\otimes k}\otimes\id)(\psi),
\quad 1\le k\le n,
\end{align}
where the identity acts on the untouched input systems.
Including these systems in the reference gives
\begin{align}
D(\rho_k\|\sigma_k)
\le D(\N^{\otimes k}\|\M^{\otimes k})
\le kD^{\mathrm{reg}}(\N\|\M)<\infty.
\end{align}
For each $k=1,\ldots,n-1$, let $B_1,\ldots,B_k$
denote the outputs of the first $k$ channel uses and set
$S_k:=RB_1\cdots B_kA_{k+2}\cdots A_n$.
After reordering tensor factors, $\rho_k$ and $\sigma_k$
are states on $S_k\otimes A_{k+1}$, and
\begin{align}
D(\rho_{k+1}\|\sigma_{k+1})
&=D\bigl((\id_{S_k}\otimes\N)(\rho_k)
\big\|(\id_{S_k}\otimes\M)(\sigma_k)\bigr)\le D(\rho_k\|\sigma_k)+D^{\mathrm{reg}}(\N\|\M).
\end{align}
The equality applies the next channel use to $A_{k+1}$,
leaving $S_k$ unchanged.
The inequality is the chain rule \cite[Corollary~3]{FFRS}
applied to $\rho_k,\sigma_k$ with reference system $S_k$;
their relative entropy is finite by the preceding bound.
Taking the supremum over all input states $\psi_{RA_1\cdots A_n}$ gives
\begin{align}\label{eq:sequential-upper}
D(\N^{\otimes n}\|\M^{\otimes n})
&=\sup_{\psi}D(\rho_n\|\sigma_n)\notag\\
&=\sup_{\psi}\Bigl\{D(\rho_1\|\sigma_1)
+\sum_{k=1}^{n-1}\bigl[D(\rho_{k+1}\|\sigma_{k+1})
-D(\rho_k\|\sigma_k)\bigr]\Bigr\}\notag\\
&\le\sup_{\psi}D(\rho_1\|\sigma_1)
+(n-1)D^{\mathrm{reg}}(\N\|\M)\notag\\
&\le D(\N\|\M)+(n-1)D^{\mathrm{reg}}(\N\|\M).
\end{align}
The first equality is the definition of channel relative entropy,
and the second telescopes the relative-entropy increments.
The first inequality sums the chain-rule bounds over $k=1,\ldots,n-1$.
For the last inequality, regard $RA_2\cdots A_n$ as the reference
for the first channel use. The resulting output divergence is
at most $D(\N\|\M)$ for every $R$ and $\psi$.
Rearranging \eqref{eq:sequential-upper}, and including $n=1$
where equality holds, gives
\begin{align}\label{eq:universal-gap}
D^{\mathrm{reg}}(\N\|\M)
-\frac1nD(\N^{\otimes n}\|\M^{\otimes n})
\ge
\frac{D^{\mathrm{reg}}(\N\|\M)-D(\N\|\M)}{n},
\qquad n\ge1.
\end{align}
If some finite block attains $D^{\mathrm{reg}}(\N\|\M)$,
this inequality forces
$D^{\mathrm{reg}}(\N\|\M)=D(\N\|\M)$.
Conversely, superadditivity implies that attainment at
one copy yields attainment at every blocklength.
Thus exactly one of the two alternatives holds.
\end{proof}

\begin{boxremark}[Lower bound on finite-block approximation]
\label{rem:block-lower-bound}
Under the assumptions of Theorem~\ref{thm:universal}, suppose
$D^{\mathrm{reg}}(\N\|\M)>D(\N\|\M)$.
For $\varepsilon>0$, \eqref{eq:universal-gap} implies that any
blocklength $m$ satisfying
\begin{align}
D^{\mathrm{reg}}(\N\|\M)
-\frac1m D(\N^{\otimes m}\|\M^{\otimes m})
\le\varepsilon
\end{align}
must obey
$m\ge [D^{\mathrm{reg}}(\N\|\M)-D(\N\|\M)]/\varepsilon$.
Thus finite-block approximation requires
$m=\Omega(\varepsilon^{-1})$ for every fixed
nonadditive channel pair. Under the additional assumption
$J_{\M}>0$, Theorem~\ref{thm:physical-block} in Section~\ref{sec:regularized-computation} gives the complementary
sufficient blocklength $\widetilde O(\varepsilon^{-3/2})$.
\end{boxremark}

\begin{boxremark}[General divergences]
The dichotomy can be generalized to other divergences. For a state divergence $\mathbb D$, define the channel divergence
$\mathbb D(\N\|\M)$ by replacing $D$ with $\mathbb D$ in
\eqref{eq:channel-relative-entropy}, and set
\begin{align}
\mathbb D^{\mathrm{reg}}(\N\|\M)
:=\sup_{m\ge1}\frac1m
\mathbb D(\N^{\otimes m}\|\M^{\otimes m}).
\end{align}
The same dichotomy as in Theorem~\ref{thm:universal} holds under the following conditions:
\begin{enumerate}[label=\textnormal{(\roman*)}]
\item \emph{Finiteness.}
$\mathbb D^{\mathrm{reg}}(\N\|\M)<\infty$.
\item \emph{Superadditivity.} For all $m,n\ge1$,
\begin{align}
\mathbb D(\N^{\otimes(m+n)}\|\M^{\otimes(m+n)})
\ge
\mathbb D(\N^{\otimes m}\|\M^{\otimes m})
+\mathbb D(\N^{\otimes n}\|\M^{\otimes n}).
\end{align}
\item \emph{Chain rule.}
For every pair of states
$\rho,\sigma\in\mathcal D(R\otimes A)$ with
$\mathbb D(\rho\|\sigma)<\infty$,
\begin{align}
\mathbb D\bigl((\id_R\otimes\N)(\rho)
\big\|(\id_R\otimes\M)(\sigma)\bigr)
\le \mathbb D(\rho\|\sigma)
+\mathbb D^{\mathrm{reg}}(\N\|\M).
\end{align}
\end{enumerate}
In particular, the sandwiched R\'enyi divergences
$\mathbb D=\widetilde D_\alpha$, $1<\alpha<\infty$,
defined in Equation~\eqref{eq:s7-sandwiched-renyi}, satisfy these conditions when
$\supp J_{\N}\subseteq\supp J_{\M}$.
Indeed, finiteness follows from Choi domination,
superadditivity from product-state additivity, and the
chain rule from \cite[Corollary~5.2]{FF21}.
\end{boxremark}

\section{Finite-attainment criterion for interior optimizers}
\label{sec:interior-modular}

The finite-attainment dichotomy shows that the regularized channel relative entropy is either attained with one copy or never attained at any finite blocklength. Distinguishing these cases by directly comparing $D^{\mathrm{reg}}(\N\|\M)$ with $D(\N\|\M)$ is difficult as the regularized divergence involves an asymptotic optimization. Here, we show that two-copy additivity provides an equivalent criterion whenever a one-copy optimal pure input has a full-rank reference marginal (dubbed as \emph{interior optimizer}). This full-rank condition holds generically, as established in Section~\ref{sec:generic}, so the criterion applies to almost all channel pairs.

\begin{boxtheorem}[Finite attainment with an interior optimizer]
\label{thm:modular}
Let $A, B$ be finite-dimensional systems, and $\N,\M$ be channels from $A$ to $B$, with $D(\N\|\M)<\infty$.
Suppose a one-copy optimal pure input $\psi_{RA}$, with $R\simeq A$,
has full-rank reference marginal $\tau=\Tr_A\psi_{RA}$, and write
\begin{align}
\rho=(\id_R\otimes\N)(\psi),
\qquad
\sigma=(\id_R\otimes\M)(\psi).
\end{align}
The following three conditions are equivalent:
\begin{enumerate}[
    label=\textnormal{(\roman*)},
    ref=(\roman*)
]
\item\label{item:modular-reg} \emph{One-copy attainment}:
$D^{\mathrm{reg}}(\N\|\M)=D(\N\|\M)$.

\item\label{item:modular-two} \emph{Two-copy additivity}:
$D(\N^{\otimes 2}\|\M^{\otimes 2})=2D(\N\|\M)$.

\item\label{item:modular-flow} \emph{Modular-flow invariance}:
$\operatorname{Tr}_B(\sigma^{it}\rho\sigma^{-it})=\tau$
for every $t\in\mathbb R$, with $\sigma^{it}$
defined on $\supp\sigma$.
\end{enumerate}
\end{boxtheorem}

The proof of these equivalences is given in Section~\ref{sec:interior-proof} and follows the cycle
\ref{item:modular-two}~$\Rightarrow$
\ref{item:modular-flow}~$\Rightarrow$
\ref{item:modular-reg}~$\Rightarrow$ \ref{item:modular-two}.
First, two-copy additivity makes $\psi_{RA}^{\otimes2}$ optimal.
Stationarity at one and two copies gives first-order optimality
conditions expressed through the Bregman remainder. The two-copy
identity in Lemma~\ref{lem:covariance} then forces the reference
marginal to remain unchanged under the modular flow of $\sigma$.
Next, this invariance extends the one-copy optimality condition to
product inputs at every blocklength (Lemma~\ref{lem:tensorize}).
The supporting bound in Lemma~\ref{lem:supporting-bound} shows that
these product inputs are globally optimal, establishing additivity
at every blocklength and hence one-copy attainment.
Finally, \ref{item:modular-reg}~$\Rightarrow$ \ref{item:modular-two}
follows immediately from the superadditivity of channel relative entropy.

\subsection{Bregman bounds for relative entropy}

We first define the Bregman remainder, the first-order Taylor
remainder of $f(x)=x\ln x$. Its trace equals relative entropy,
and it is used below to compute the derivative needed for the
supporting bound.

\begin{boxdefinition}[Bregman remainder]\label{def:Bregman}
Let $\mathcal H$ be a finite-dimensional Hilbert space,
and let $X,Y\in\mathcal B(\mathcal H)$ be positive semidefinite with $\supp X\subseteq\supp Y$.
Define the Bregman remainder by
\begin{align}
\mathfrak B(X,Y)
:=
f(X)-f(Y)-\mathfrak Df_Y(X-Y),
\end{align}
where $f(x)=x\ln x$, and $\mathfrak Df_Y$ denotes the Fr\'echet derivative of $f$ at $Y$,
taken on $\supp Y$. The resulting operator is extended by zero on
$(\supp Y)^\perp$.
\end{boxdefinition}

Following the Bregman operator divergence
\cite{Petz07}, the standard trace and
Fr\'echet derivative formulas
\cite[Theorem~3.23 and Example~3.34]{HP14} give
\begin{align}
\Tr\mathfrak Df_Y(Z)&=\Tr Z(\ln Y+I),
\label{eq:trace-frechet}\\
\mathfrak B(X,Y)&=f(X)+Y-\mathfrak Df_Y(X),
\label{eq:bregman-form}\\
\Tr\mathfrak B(X,Y)&=D(X\|Y),
\label{eq:bregman-trace}
\end{align}
where $I$ is the identity on $\supp Y$ and $Z$
is any operator supported on $\supp Y$.

The next lemma shows that
$D(\sqrt H\,X\sqrt H\|\sqrt H\,Y\sqrt H)$ is concave in $H$.

\begin{boxlemma}[Concavity]\label{lem:filtered-concavity}
Let $\mathcal H$ be a finite-dimensional Hilbert space,
and let $X,Y\in\mathcal B(\mathcal H)$ be positive semidefinite with
$\supp X\subseteq\supp Y$.
For a positive semidefinite $H\in\mathcal B(\mathcal H)$, define
\begin{align}
D_{X,Y}(H):=D(\sqrt H\,X\sqrt H\|\sqrt H\,Y\sqrt H).
\end{align}
Then $D_{X,Y}$ is concave on the positive semidefinite cone.
\end{boxlemma}

\begin{proof}
Let $H,H_1,H_2\ge0$. We first establish
\begin{align}\label{eq:q-superadditivity}
D_{X,Y}(H_1+H_2)\ge D_{X,Y}(H_1)+D_{X,Y}(H_2),
\qquad D_{X,Y}(tH)=tD_{X,Y}(H)\quad(t\ge0).
\end{align}
For the first inequality, define the linear operator
$L:\mathcal H\to\mathcal H\oplus\mathcal H$ by
$Lv=\sqrt{H_1}v\oplus\sqrt{H_2}v$, so that
$L^\dagger L=H_1+H_2$. Its polar decomposition is
$L=V\sqrt{H_1+H_2}$, where $V$ is isometric on
$\supp(H_1+H_2)$. Let $\mathcal P$ denote the pinching channel
that removes the off-diagonal blocks between the two summands.
Then we have
$D_{X,Y}(H_1+H_2)
=D(LXL^\dagger\|LYL^\dagger)
\ge D\bigl(\mathcal P(LXL^\dagger)\big\|
\mathcal P(LYL^\dagger)\bigr)
=D_{X,Y}(H_1)+D_{X,Y}(H_2)$.
The first equality uses invariance under the common isometry $V$,
and the inequality is data processing for $\mathcal P$.
The last equality follows from additivity of relative entropy over
direct sums, whose blocks are $\sqrt{H_i}X\sqrt{H_i}$ and
$\sqrt{H_i}Y\sqrt{H_i}$ for $i=1,2$.
For $t>0$, the identity $\sqrt{tH}=\sqrt t\,\sqrt H$ and
$D(tA\|tB)=tD(A\|B)$ give $D_{X,Y}(tH)=tD_{X,Y}(H)$; the case $t=0$
follows from $D_{X,Y}(0)=0$.

Combining these two properties, for every $0\le\lambda\le1$, we get
$D_{X,Y}\bigl(\lambda H_1+(1-\lambda)H_2\bigr)
\ge D_{X,Y}(\lambda H_1)+D_{X,Y}\bigl((1-\lambda)H_2\bigr)
=\lambda D_{X,Y}(H_1)+(1-\lambda)D_{X,Y}(H_2)$.
Thus $D_{X,Y}$ is concave.
\end{proof}

We then express the derivative of $D_{X,Y}$ at $H=I$ using the Bregman remainder
and obtain a supporting bound for arbitrary linear operators.

\begin{boxlemma}[Supporting bounds]\label{lem:supporting-bound}
Let $\mathcal H$ be a finite-dimensional Hilbert space,
and let $X,Y\in\mathcal B(\mathcal H)$ be positive semidefinite with
$\supp X\subseteq\supp Y$.
For $D_{X,Y}$ defined in Lemma~\ref{lem:filtered-concavity}
and every Hermitian $K\in\mathcal B(\mathcal H)$,
\begin{align}
\left.\frac{d}{ds}\right|_{s=0}D_{X,Y}(I+sK)
=\Tr K\,\mathfrak B(X,Y).
\end{align}
Here $I$ is the identity on $\mathcal H$ and $\mathfrak B(X,Y)$ is
the Bregman remainder from Definition~\ref{def:Bregman}.
The path $I+sK$ remains positive definite for all sufficiently small
$|s|$. Consequently, for every
linear operator $L:\mathcal H\to\mathcal K$ to a finite-dimensional Hilbert
space $\mathcal K$,
\begin{align}
D(LXL^\dagger\|LYL^\dagger)
\le \Tr L^\dagger L\,\mathfrak B(X,Y),
\qquad
\mathfrak B(X,Y)\ge0.
\end{align}
\end{boxlemma}

\begin{proof}
Let $P$ be the orthogonal projection onto
$\supp Y$. The polar decomposition of $\sqrt H\,P$ and invariance of
relative entropy under a common isometry show that $D_{X,Y}(H)$ depends only
on $PHP$. We may therefore work on $\supp Y$, where $Y$ is positive
definite, and replace $K$ by its compression $PKP$.

For Hermitian $K$, let $H_s=I+sK$, and set
$X_s=\sqrt{H_s}\,X\sqrt{H_s},
Y_s=\sqrt{H_s}\,Y\sqrt{H_s}.$
We use $\sqrt{H_s}=I+\frac{s}{2}K+O(s^2)$
and write $\{Z,W\}:=ZW+WZ$.
Then \eqref{eq:trace-frechet} and the product
rule give
\begin{align}
\left.\frac{d}{ds}\right|_{s=0}D_{X,Y}(H_s)
&=
\left.\frac{d}{ds}\right|_{s=0}
\Bigl[
\Tr X_s\ln X_s
-\Tr X_s\ln Y_s
-\Tr X_s+\Tr Y_s
\Bigr]\notag\\
&=
\Tr\frac12\{K,X\}(\ln X+I)
-\Tr\frac12\{K,X\}\ln Y\notag\\
&\quad
-\Tr X\,\mathfrak D\ln_Y
   \!\left(\frac12\{K,Y\}\right)
-\Tr\frac12\{K,X\}
+\Tr\frac12\{K,Y\}\notag\\
&=
\Tr\frac12\{K,X\}(\ln X-\ln Y)
-\Tr X\,\mathfrak D\ln_Y
   \!\left(\frac12\{K,Y\}\right)
+\Tr\frac12\{K,Y\}\notag\\
&=
\Tr K\!\left[
X\ln X-\frac12\{X,\ln Y\}
+Y-\frac12\{Y,\mathfrak D\ln_Y(X)\}
\right]\notag\\
&=\Tr K\,\mathfrak B(X,Y).
\end{align}
If $X$ is singular, the term $\Tr X_s\ln X_s$ is differentiated
on the nonzero spectrum of $X_s$, whose rank is constant for small $|s|$.
The logarithm of $X$ is taken on $\supp X$ and extended by zero
on its orthogonal complement. The fourth equality uses the self-adjointness
of $\mathfrak D\ln_Y$ under the trace inner product,
while the last follows from
\eqref{eq:bregman-form} and the product rules for
$Y\ln Y=\ln Y\,Y$.

By \eqref{eq:bregman-trace},
$D_{X,Y}(I)=D(X\|Y)=\Tr\mathfrak B(X,Y)$.
By Lemma~\ref{lem:filtered-concavity}, concavity yields
\begin{align}
D_{X,Y}(H)
\le D_{X,Y}(I)+\mathfrak D(D_{X,Y})_I(H-I)
=\Tr H\,\mathfrak B(X,Y).
\end{align}
For $L=U\sqrt{L^\dagger L}$, taking
$H=L^\dagger L$ and using invariance
of $D$ under a common partial isometry gives
\begin{align}
D(LXL^\dagger\|LYL^\dagger)
=D_{X,Y}(H)
\le \Tr L^\dagger L\,\mathfrak B(X,Y).
\end{align}

Finally, Klein's inequality gives
$D_{X,Y}(H)\ge0$. Therefore,
$\Tr H\,\mathfrak B(X,Y)\ge0$
for every $H\ge0$.
Taking $H=|v\rangle\langle v|$
yields $\mathfrak B(X,Y)\ge0$.
\end{proof}

\Needspace{10\baselineskip}
\subsection{Modular flow and tensorization}

We next compare the two-copy Bregman term with the sum of its
one-copy contributions. The next lemma writes their difference
as an integral representation and shows that it vanishes exactly when the
modular-flow condition holds.

\begin{boxlemma}[Two-copy modular identity]
\label{lem:covariance}
Let $\mathcal H,\mathcal K$ be finite-dimensional Hilbert spaces,
and let $X,Y\in\mathcal B(\mathcal H)$ be positive semidefinite with $\supp X\subseteq\supp Y$.
Let $\Gamma:\mathcal B(\mathcal H)\to\mathcal B(\mathcal K)$
be a Hermiticity-preserving linear map with $\Gamma(X)=\Gamma(Y)=:\omega$. Define
\begin{align}
T_m&:=\Gamma^{\otimes m}
\mathfrak B(X^{\otimes m},Y^{\otimes m}),\qquad m\ge1,\\
\Delta_t&:=\Gamma(Y^{it}XY^{-it})-\omega,\qquad t\in\mathbb R,
\end{align}
where $\mathfrak B$ is the Bregman remainder defined in Definition~\ref{def:Bregman} and
$Y^{it}$ is defined on $\supp Y$. Then
\begin{align}
T_2-T_1\otimes\omega-\omega\otimes T_1
=
\int_{\mathbb R}
\frac{\pi}{2\sinh^2(\pi t)}
\,\Delta_t\otimes\Delta_t\,dt.
\end{align}
Hence the left-hand side vanishes if and only if
$\Delta_t=0$ for every $t$.
\end{boxlemma}

\begin{proof}
All spectral calculations below are performed on $\supp Y$, where
$Y$ is positive definite, and the resulting operators are extended by
zero on $(\supp Y)^\perp$.
Using the product rules for $Y\ln Y=\ln Y\,Y$, decompose
the Bregman remainder $\mathfrak B(X,Y)$ into $U(X,Y)+Y-M_Y(X)$, where
\begin{align}
U(X,Y):=X\ln X-\frac12(X\ln Y+\ln Y\,X),
\\
M_Y(X):=\frac12\bigl(Y\,\mathfrak D\ln_Y(X)+\mathfrak D\ln_Y(X)Y\bigr).
\end{align}

Diagonalize $Y=\sum_i y_i|i\rangle\langle i|$ on $\supp Y$, so $y_i>0$. By the
Fr\'echet derivative formula \cite[Example~3.34]{HP14},
\begin{align}
\bigl(\mathfrak D\ln_Y(X)\bigr)_{ij}
=
\begin{cases}
\dfrac{\ln y_i-\ln y_j}{y_i-y_j}\,X_{ij},&y_i\ne y_j,\\[1ex]
\dfrac1{y_i}X_{ij},&y_i=y_j.
\end{cases}
\end{align}
Set $u:=\ln y_i-\ln y_j$, so that
$y_i=\sqrt{y_iy_j}e^{u/2}$ and $y_j=\sqrt{y_iy_j}e^{-u/2}$. For $u\ne0$,
\begin{align}
(M_Y(X))_{ij}
=
\frac{y_i+y_j}{2}
\frac{\ln y_i-\ln y_j}{y_i-y_j}\,X_{ij}
=
\frac u2
\frac{e^{u/2}+e^{-u/2}}{e^{u/2}-e^{-u/2}}\,X_{ij}
=
\frac u2\coth\frac u2\,X_{ij},
\end{align}
while for $u=0$ both sides equal $X_{ij}$ by continuity.
Since
\begin{align}
\frac u2\coth\frac u2-1
=
\int_{\mathbb R}\frac{\pi}{2\sinh^2(\pi t)}(1-\cos ut)\,dt,
\end{align}
and $(Y^{it}XY^{-it})_{ij}=e^{iut}X_{ij}$, we obtain
\begin{align}
(M_Y(X))_{ij}
&=
X_{ij}+
\int_{\mathbb R}
\frac{\pi}{2\sinh^2(\pi t)}
(1-\cos ut)X_{ij}\,dt\notag\\
&=
\left[
X+
\int_{\mathbb R}
\frac{\pi}{2\sinh^2(\pi t)}
\left(
X-\frac{Y^{it}XY^{-it}+Y^{-it}XY^{it}}2
\right)dt
\right]_{ij}.
\end{align}
Moreover, 
using $\ln(x\otimes x)=\ln x\otimes I+I\otimes\ln x$,
we have
\begin{align}
U(X\otimes X,Y\otimes Y)
=U(X,Y)\otimes X+X\otimes U(X,Y).
\end{align}

Since $\Gamma(X)=\Gamma(Y)=\omega$ and
$\Gamma(Y^{it}XY^{-it})=\omega+\Delta_t$, we obtain
\begin{align}
T_1
&=
\Gamma U(X,Y)
+\int_{\mathbb R}
\frac{\pi}{2\sinh^2(\pi t)}\,
\frac{\Delta_t+\Delta_{-t}}2\,dt,
\end{align}
and
\begin{align}
T_2
&=\Gamma U(X,Y)\otimes\omega+\omega\otimes\Gamma U(X,Y)\notag\\
&\quad+
\int_{\mathbb R}
\frac{\pi}{2\sinh^2(\pi t)}
\left[
\frac{(\omega+\Delta_t)^{\otimes2}+(\omega+\Delta_{-t})^{\otimes2}}2
-\omega^{\otimes2}
\right]dt.
\end{align}
The symmetric numerators are $O(t^2)$ at the origin, so these
integrals converge absolutely. Subtracting and using the evenness of
$\pi/[2\sinh^2(\pi t)]$ yields
\begin{align}
T_2-T_1\otimes\omega-\omega\otimes T_1
=
\int_{\mathbb R}
\frac{\pi}{2\sinh^2(\pi t)}
\,\Delta_t\otimes\Delta_t\,dt.
\end{align}

Let $F$ be the swap operator on $\mathcal K\otimes\mathcal K$.
For operators $Z,W$ on $\mathcal K$, one has
$\operatorname{Tr}F(Z\otimes W)=\operatorname{Tr}(ZW)$. Hence
\begin{align}
\operatorname{Tr}F
\bigl(T_2-T_1\otimes\omega-\omega\otimes T_1\bigr)
=
\int_{\mathbb R}
\frac{\pi}{2\sinh^2(\pi t)}
\operatorname{Tr}F(\Delta_t\otimes\Delta_t)\,dt
=
\int_{\mathbb R}
\frac{\pi}{2\sinh^2(\pi t)}
\|\Delta_t\|_2^2\,dt.
\end{align}
Since the integrand is nonnegative, vanishing of the left-hand side implies
$\Delta_t=0$ almost everywhere. By continuity of $t\mapsto\Delta_t$, this
holds for every $t$. The converse is immediate.
\end{proof}

The next lemma proves that the modular-flow invariance condition extends to every blocklength, so that the Bregman remainder tensorizes. This is the key step in proving that one-copy optimality implies optimality at every blocklength.

\begin{boxlemma}[Tensorization under modular invariance]
\label{lem:tensorize}
Let $\mathcal H,\mathcal K$ be finite-dimensional Hilbert spaces,
and let $X,Y\in\mathcal B(\mathcal H)$ be positive semidefinite with $\supp X\subseteq\supp Y$.
Let $\Gamma:\mathcal B(\mathcal H)\to\mathcal B(\mathcal K)$
be a Hermiticity-preserving linear map with $\Gamma(X)=\Gamma(Y)=:\omega$.
Suppose that
$\Gamma(Y^{it}XY^{-it})=\omega$
for every $t\in\mathbb R$.
Then, with $T_m$ as in Lemma~\ref{lem:covariance}, it holds that
\begin{align}
T_m
=
\sum_{j=1}^m
\omega^{\otimes(j-1)}
\otimes T_1
\otimes\omega^{\otimes(m-j)},
\qquad m\ge1.
\end{align}
\end{boxlemma}

\begin{proof}
By the tensor-product rule for logarithms on the respective supports,
\begin{align}
U(X^{\otimes m},Y^{\otimes m})
=
\sum_{j=1}^m
X^{\otimes(j-1)}\otimes U(X,Y)\otimes X^{\otimes(m-j)}.
\end{align}
Moreover, on $(\supp Y)^{\otimes m}$,
$(Y^{\otimes m})^{it}=(Y^{it})^{\otimes m}$ since
$\ln(Y^{\otimes m})
=\sum_{j=1}^m I^{\otimes(j-1)}\otimes\ln Y\otimes I^{\otimes(m-j)}$,
where $I$ is the identity on $\supp Y$.

Using the integral representation of $M_Y$ from the proof of
Lemma~\ref{lem:covariance} and
$\Gamma(Y^{it}XY^{-it})=\omega$, we obtain
\begin{align}
T_m
&=
\Gamma^{\otimes m}U(X^{\otimes m},Y^{\otimes m})
+\omega^{\otimes m}
-\Gamma^{\otimes m}M_{Y^{\otimes m}}(X^{\otimes m})\notag\\
&=
\sum_{j=1}^m
\omega^{\otimes(j-1)}\otimes\Gamma U(X,Y)\otimes\omega^{\otimes(m-j)}\notag\\
&\quad-
\int_{\mathbb R}
\frac{\pi}{2\sinh^2(\pi t)}
\left[
\omega^{\otimes m}
-\frac{\Gamma(Y^{it}XY^{-it})^{\otimes m}
+\Gamma(Y^{-it}XY^{it})^{\otimes m}}2
\right]dt\notag\\
&=
\sum_{j=1}^m
\omega^{\otimes(j-1)}\otimes T_1\otimes\omega^{\otimes(m-j)},
\end{align}
For $m=1$, the identity
$\Gamma M_Y(X)=\omega=\Gamma(Y)$ gives $T_1=\Gamma U(X,Y)$.
\end{proof}

\subsection{Proof of the finite-attainment criterion for interior optimizers}
\label{sec:interior-proof}

With the supporting bounds in Lemma~\ref{lem:supporting-bound}, the two-copy modular identity in Lemma~\ref{lem:covariance}, and the tensorization property in Lemma~\ref{lem:tensorize}, we are now ready to prove Theorem~\ref{thm:modular}. 

\begin{proof}[Proof of Theorem~\ref{thm:modular}]
In the preceding lemmas, take
$(X,Y,\Gamma,\omega)=
(\rho,\sigma,\operatorname{Tr}_B,\tau)$,
restricting to $\supp\sigma$ if necessary, with $\Gamma$ understood as
the embedding into $RB$ followed by $\operatorname{Tr}_B$.

For any Hermitian $K$ satisfying $\operatorname{Tr}\tau K=0$, set
$H_s=I+sK$. Then $\operatorname{Tr}H_s\tau=1$, and $H_s>0$ for all
sufficiently small $|s|$. Hence optimality of $\psi$ gives
\begin{align}
0
=
\left.\frac{d}{ds}\right|_{s=0}D_{\rho,\sigma}(H_s\otimes I_B)
=
\operatorname{Tr}(K\otimes I_B)\mathfrak B(\rho,\sigma)
=
\operatorname{Tr}KT_1,
\end{align}
where the second equality follows from Lemma~\ref{lem:supporting-bound}. Thus
$\operatorname{Tr}KT_1=0$
whenever
$\operatorname{Tr}\tau K=0$.
Hence $T_1$ is orthogonal to every Hermitian operator orthogonal to
$\tau$, so $T_1=\lambda\tau$ for some $\lambda$. Since
$\operatorname{Tr}\tau=1$, Equation~\eqref{eq:bregman-trace} gives
\begin{align}
\lambda
=
\operatorname{Tr}T_1
=
\operatorname{Tr}\mathfrak B(\rho,\sigma)
=
D(\rho\|\sigma)
=
D(\N\|\M).
\end{align}
Therefore $T_1=D(\N\|\M)\tau$.

\medskip
\textbf{\emph{\ref{item:modular-two}
$\Rightarrow$ \ref{item:modular-flow}.}}
If $D(\N^{\otimes 2}\|\M^{\otimes 2})=2D(\N\|\M)$, then $\psi^{\otimes2}$ is optimal, and the same
stationarity argument gives $T_2=2D(\N\|\M)\,\tau^{\otimes2}$. Since
$T_1=D(\N\|\M)\tau$,
\begin{align}
T_2-T_1\otimes\tau-\tau\otimes T_1=0,
\end{align}
so Lemma~\ref{lem:covariance} yields
$\operatorname{Tr}_B(\sigma^{it}\rho\sigma^{-it})=\tau$ for all $t$.

\medskip
\textbf{\emph{\ref{item:modular-flow}
$\Rightarrow$ \ref{item:modular-reg}.}}
Now suppose
$\operatorname{Tr}_B(\sigma^{it}\rho\sigma^{-it})=\tau$ for all $t$.
Lemma~\ref{lem:tensorize} and $T_1=D(\N\|\M)\tau$ give
$T_m=mD(\N\|\M)\,\tau^{\otimes m}$. By the pure-input reduction, it suffices
to consider pure $m$-copy inputs. Since $\tau>0$, $\psi^{\otimes m}$ has
full Schmidt rank, so every pure $m$-copy input is of the form
\begin{align}
(V\otimes I)\psi^{\otimes m}(V^\dagger\otimes I),
\qquad
\operatorname{Tr}V^\dagger V\,\tau^{\otimes m}=1.
\end{align}
Lemma~\ref{lem:supporting-bound} therefore gives
\begin{align}
D(\N^{\otimes m}\|\M^{\otimes m})\le\operatorname{Tr}V^\dagger V\,T_m=mD(\N\|\M).
\end{align}
Since $\frac1mD(\N^{\otimes m}\|\M^{\otimes m})\ge D(\N\|\M)$, we have
$\frac1mD(\N^{\otimes m}\|\M^{\otimes m})=D(\N\|\M)$ for all $m$, hence $D^{\mathrm{reg}}(\N\|\M)=D(\N\|\M)$.

\medskip
\textbf{\emph{\ref{item:modular-reg}
$\Rightarrow$ \ref{item:modular-two}.}}
Finally, $D^{\mathrm{reg}}(\N\|\M)=D(\N\|\M)$ implies
\begin{align}
2D(\N\|\M)\le D(\N^{\otimes 2}\|\M^{\otimes 2})
\le 2D^{\mathrm{reg}}(\N\|\M)=2D(\N\|\M),
\end{align}
so $D(\N^{\otimes 2}\|\M^{\otimes 2})=2D(\N\|\M)$.
\end{proof}

\section{Finite-attainment criterion for qubit channels}\label{sec:qubit}

The finite-attainment criterion in Theorem~\ref{thm:modular} requires a full-rank reference
marginal. In this section, we remove this assumption for qubit
channels by treating rank-one optimal reference marginals separately. 
Consequently, for every qubit channel pair with finite relative entropy,
two-copy additivity always determines whether the regularized value is attained
at a finite blocklength.
\begin{boxtheorem}[Finite attainment for qubit channels]
\label{thm:qubit}
Let $\N,\M:A\to B$ be quantum channels with $|A|=|B|=2$ and
$D(\N\|\M)<\infty$.
The following two conditions are equivalent:
\begin{enumerate}[label=\textnormal{(\roman*)},ref=(\roman*)]
\item\label{item:qubit-reg} \emph{One-copy attainment:}
$D^{\mathrm{reg}}(\N\|\M)=D(\N\|\M)$.
\item\label{item:qubit-two} \emph{Two-copy additivity:}
$D(\N^{\otimes 2}\|\M^{\otimes 2})=2D(\N\|\M)$.
\end{enumerate}
\end{boxtheorem}

The proof of this equivalence is given in
Section~\ref{sec:qubit-criterion-proof}. For
\ref{item:qubit-two}~$\Rightarrow$ \ref{item:qubit-reg},
the interior case follows from Theorem~\ref{thm:modular}, leaving
the rank-one case to be considered. First, boundary perturbations
give alignment and optimality conditions for the Choi blocks.
Combined with the qubit trace constraint and two-copy additivity,
these conditions force either commuting output states at the optimizer
or vanishing off-diagonal blocks (Lemma~\ref{lem:qubit-boundary}).
Next, both alternatives yield decompositions with orthogonally flagged
branches. The direct-sum formula for relative entropy then gives a
chain-rule bound with the one-copy channel divergence, valid for
arbitrary reference systems. Applying this bound successively proves
additivity at every blocklength and hence one-copy attainment.
Finally, \ref{item:qubit-reg}~$\Rightarrow$ \ref{item:qubit-two}
follows immediately from the superadditivity of channel relative entropy.

\subsection{Boundary optimality and rigidity}

\begin{boxlemma}\label{lem:centering}
Let $\mathcal H$ be a two-dimensional Hilbert space.
Consider $N,M\in\mathcal D(\mathcal H)$, and
$Z\in\mathcal B(\mathcal H)$.
If $M>0$, $N\ne M$, and
$\Tr NZ=\Tr MZ=\Tr\mathfrak B(N,M)Z=0$,
then we have $\Tr Z=0$.
\end{boxlemma}

\begin{proof}
For a qubit state $T$, the resolvent representation of $f(x)=x\ln x$ gives
\begin{align}
f(T)
&=\int_0^\infty\left(\frac{T}{1+s}-T(T+sI)^{-1}\right)ds
=\alpha(T)T+\beta(T)I,\notag\\
\beta(T)&:=-\int_0^\infty\frac{\det T}{s^2+s+\det T}\,ds.
\end{align}
The second equality follows from Cayley--Hamilton, which gives
$(T+sI)^{-1}=[(1+s)I-T]/(s^2+s+\det T)$ for $s>0$;
$\alpha(T)$ is a real scalar. These formulas include singular $T$ by continuity.

On qubit states, $\det T=(1-\Tr T^2)/2$ is strictly concave.
For every $s>0$, the scalar function $x\mapsto-x/(s^2+s+x)$
is strictly decreasing and convex on $[0,\infty)$.
Thus $\beta$ is strictly convex on qubit states and differentiable at $M>0$.

Set $T_t=M+t(N-M)$. By the definition of the Bregman remainder,
\begin{align}
\Tr\mathfrak B(N,M)Z
&=\Tr f(N)Z-\Tr f(M)Z
-\left.\frac{d}{dt}\Tr f(T_t)Z\right|_{t=0}\notag\\
&=\left[\beta(N)-\beta(M)
-\left.\frac{d}{dt}\beta(T_t)\right|_{t=0}\right]\Tr Z.
\end{align}
Here we used $\Tr T_tZ=(1-t)\Tr MZ+t\Tr NZ=0$.
Strict convexity and $N\ne M$ make the coefficient of $\Tr Z$ strictly positive.
The assumed vanishing of the left-hand side therefore implies $\Tr Z=0$.
\end{proof}

The next lemma shows that two-copy additivity imposes a simple
structural alternative at a rank-one optimizer. Either the optimal
output states commute or both off-diagonal Choi blocks vanish.
This rigidity is the structural ingredient in the chain-rule argument below.

\begin{boxlemma}[Boundary rigidity]
\label{lem:qubit-boundary}
Let $\N,\M$ be qubit channels with
$0<D(\N\|\M)<\infty$, and suppose a rank-one state $\tau_0$ maximizes
the one-copy objective in \eqref{eq:input-objective}.
In a reference basis with $\tau_0=|0\rangle\langle0|$, write
\begin{align}
J_{\N}&=\begin{pmatrix}N_0&X\\X^\dagger&N_1\end{pmatrix},
\qquad
J_{\M}=\begin{pmatrix}M_0&Y\\Y^\dagger&M_1\end{pmatrix}.
\end{align}
If $D(\N^{\otimes2}\|\M^{\otimes2})=2D(\N\|\M)$,
then $[N_0,M_0]=0$ or $X=Y=0$.
\end{boxlemma}

\begin{proof}
To keep the main argument clear, we defer the detailed derivations of
the boundary perturbation estimates to Sections~\ref{app:alignment}
and~\ref{app:inward}. We use these estimates below to derive
block alignment and a coherent trace condition. We then combine
these conditions with the qubit trace constraint and two-copy
additivity to prove the stated alternative.

First, $D(\N\|\M)=D(N_0\|M_0)>0$ implies $N_0\ne M_0$.
We also have $M_0>0$: otherwise support inclusion and
$\Tr N_0=\Tr M_0=1$ would imply $N_0=M_0$.
Trace preservation gives
$\Tr N_0=\Tr M_0=\Tr N_1=\Tr M_1=1$ and
$\Tr X=\Tr Y=0$.
Choose $\lambda>0$ such that $J_{\N}\le\lambda J_{\M}$.
Set $Z:=M_0^{-1}Y$, $F:=N_1-Z^\dagger N_0Z$, and
$G:=M_1-Z^\dagger M_0Z$.

\emph{Alignment.}
Take
$\tau_p=\diag(1-p,p)$. The boundary
estimate in Section~\ref{app:alignment} yields
\begin{align}
D(\N,\M;\tau_p)&\ge D(\N\|\M)
+p\ln\frac1p\,\Tr(N_0^+X-Z)^\dagger N_0(N_0^+X-Z)-O(p).
\end{align}
Optimality forces $N_0(N_0^+X-Z)=0$. Since positive semidefiniteness
of $J_{\N}$ gives $\ran X\subseteq\ran N_0$,
we have $N_0N_0^+X=X=N_0Z$. By definition, $Y=M_0Z$.
Together with the Schur complement relations, this gives
\begin{align}\label{eq:qubit-block-alignment}
X=N_0Z,\qquad Y=M_0Z,\qquad F,G\ge0,\qquad F\le\lambda G.
\end{align}
The last inequality follows by compressing
$J_{\N}\le\lambda J_{\M}$ with $\binom{-Z}{I}$.

\emph{Coherent variation.}
Perturb the rank-one marginal in a complex direction:
\begin{align}
\tau_t&=\frac1{1+|t|^2}
\begin{pmatrix}1&\bar t\\t&|t|^2\end{pmatrix}.
\end{align}
The expansion from Section~\ref{app:inward} is
\begin{align}
D(\N,\M;\tau_t)&=D(\N\|\M)
+2\Re\bigl(t\Tr\mathfrak B(N_0,M_0)Z\bigr)
+O(|t|^2|\ln|t||).
\end{align}
Optimality for every phase of $t$ yields the coherent trace condition
\begin{align}\label{eq:qubit-coherent-trace}
\Tr\mathfrak B(N_0,M_0)Z=0.
\end{align}

\emph{Rigidity.}
Assume two-copy equality. If $[N_0,M_0]=0$, there is nothing to prove.
Suppose $[N_0,M_0]\ne0$; we show that $Z=0$.
We first reduce the common factor to a single matrix direction.
Equations~\eqref{eq:qubit-block-alignment}
and~\eqref{eq:qubit-coherent-trace}, together with trace preservation
and Lemma~\ref{lem:centering}, give
\begin{align}
\Tr N_0Z&=\Tr M_0Z=\Tr Z=0.
\end{align}
Thus $Z$ is trace-orthogonal to $I,N_0,M_0$.
If $\alpha I+\beta N_0+\gamma M_0=0$,
taking commutators with $M_0$ and $N_0$ gives
$\beta[N_0,M_0]=\gamma[M_0,N_0]=0$,
hence $\alpha=\beta=\gamma=0$.
Thus $I,N_0,M_0$ are linearly independent,
and their trace-orthogonal complement
in $\mathcal B(\mathbb C^2)$ is one-dimensional.

Since $M_0>0$ and $[N_0,M_0]\ne0$,
$S=i[\ln M_0,N_0]\ne0$. Moreover, $\Tr S=\Tr N_0S=\Tr M_0S=0$. Hence $S$ spans this complement, so
$Z=zS$ for some $z\in\mathbb C$.

We next use two-copy equality to eliminate this direction.
The product input $|00\rangle$ is optimal, and $X=N_0Z$ and $Y=M_0Z$ give
$X^{\otimes2}=N_0^{\otimes2}(Z\otimes Z)$, and
$Y^{\otimes2}=M_0^{\otimes2}(Z\otimes Z)$.
Apply the pure-state expansion from Section~\ref{app:inward} on
$\operatorname{span}\{|00\rangle,|11\rangle\}$, with
$N_0,M_0,Z$ replaced by $N_0^{\otimes2},M_0^{\otimes2},Z^{\otimes2}$.
Since $M_0^{\otimes2}>0$, the same expansion gives
\begin{align}
\Tr\mathfrak B(N_0^{\otimes2},M_0^{\otimes2})
(Z\otimes Z)&=0.
\end{align}

Define $\Gamma(H)=\Tr SH$.
Since $S$ is Hermitian, $\Gamma$ is
Hermiticity-preserving, and
$\Gamma(N_0)=\Gamma(M_0)=0$.
Thus $\omega=0$ in Lemma~\ref{lem:covariance},
so the terms involving $T_1$ vanish.
Using $Z=zS$, we obtain
\begin{align}
0&=
\Tr\mathfrak B(N_0^{\otimes2},M_0^{\otimes2})
(Z\otimes Z)=z^2\int_{\mathbb R}
\frac{\pi}{2\sinh^2(\pi t)}
\bigl(\Tr S M_0^{it}N_0M_0^{-it}\bigr)^2\,dt.
\end{align}

Set $g(t)=\Tr S M_0^{it}N_0M_0^{-it}$.
Since $S$ and $M_0^{it}N_0M_0^{-it}$ are Hermitian,
$g(t)$ is real. Moreover, $g(0)=0$ and
\begin{align}
g'(0)
&=i\Tr S[\ln M_0,N_0]
=-\Tr[\ln M_0,N_0]^2
=\|[\ln M_0,N_0]\|_2^2>0,
\end{align}
since $[\ln M_0,N_0]^\dagger=-[\ln M_0,N_0]\ne0$.
Thus $g(t)>0$ for sufficiently small
$t>0$, so the integral is strictly positive.
Hence $z=0$, so $Z=0$ and \eqref{eq:qubit-block-alignment}
gives $X=Y=0$.
\end{proof}

\subsection{Proof of the qubit finite-attainment criterion}
\label{sec:qubit-criterion-proof}

We are now ready to prove Theorem~\ref{thm:qubit}.

\begin{proof}[Proof of Theorem~\ref{thm:qubit}]
Assume $D(\N^{\otimes2}\|\M^{\otimes2})=2D(\N\|\M)$.
If $D(\N\|\M)=0$, then $\N=\M$ and the conclusion is immediate.
If a one-copy optimizer has a full-rank reference marginal,
Theorem~\ref{thm:modular} applies. It therefore remains to consider
$D(\N\|\M)>0$ and a rank-one optimal reference marginal
$\tau_0=|0\rangle\langle0|$.
Use the Choi block notation of Lemma~\ref{lem:qubit-boundary}.
Its proof gives $M_0>0$ and $N_0\ne M_0$, while the lemma itself
implies $X=Y=0$ or $[N_0,M_0]=0$.
For the input, use the corresponding conjugate basis, denoted by
$|0\rangle,|1\rangle$.

Fix a finite reference system $R$ and states
$\rho,\sigma\in\mathcal D(R\otimes A)$ with $D(\rho\|\sigma)<\infty$.
Set $r_j:=\Tr[(I_R\otimes|j\rangle\langle j|)\rho]$ for $j=0,1$,
so $r_0+r_1=1$.
Identity maps on $R$ are implicit until the final inequality.
In each case, we construct channels $\widetilde\N,\widetilde\M$
whose outputs include a flag register $C$, with one orthonormal
basis vector for each branch.
Tracing out $C$ recovers $\N,\M$. In both cases, we aim to show that 
\begin{align}
D\bigl((\id_R\otimes\widetilde\N)(\rho)\big\|
(\id_R\otimes\widetilde\M)(\sigma)\bigr)
-D\bigl((\id_R\otimes\widetilde\N)(\rho)\big\|
(\id_R\otimes\widetilde\N)(\sigma)\bigr)
\le D(\N\|\M).
\end{align}

\emph{Case 1: vanishing off-diagonal blocks.}
If $X=Y=0$, define, for a qubit input $\eta$,
\begin{align}
\widetilde\N(\eta)
&=\sum_{j=0}^1\langle j|\eta|j\rangle\,
|j\rangle\langle j|_C\otimes\N(|j\rangle\langle j|),\notag\\
\widetilde\M(\eta)
&=\sum_{j=0}^1\langle j|\eta|j\rangle\,
|j\rangle\langle j|_C\otimes\M(|j\rangle\langle j|).
\end{align}
The direct-sum formula gives
\begin{align}
&D(\widetilde\N(\rho)\|\widetilde\M(\sigma))
-D(\widetilde\N(\rho)\|\widetilde\N(\sigma))\notag\\
&=\sum_{j=0}^1 r_jD\bigl(\N(|j\rangle\langle j|)\big\|
\M(|j\rangle\langle j|)\bigr)
\le D(\N\|\M).
\end{align}
The inequality follows by evaluating the channel divergence at
$|j\rangle\langle j|$, which gives the relation
$D(\N(|j\rangle\langle j|)\|\M(|j\rangle\langle j|))\le D(\N\|\M)$
for $j=0,1$.

\emph{Case 2: commuting optimal outputs.}
Suppose $[N_0,M_0]=0$. Choose a common output eigenbasis and write
$N_0=\diag(n_0,n_1)$ and $M_0=\diag(m_0,m_1)$.
Set $Z:=M_0^{-1}Y$, $F:=N_1-Z^\dagger N_0Z$, and
$G:=M_1-Z^\dagger M_0Z$.
The alignment relations and residual bounds in
\eqref{eq:qubit-block-alignment} apply.
Trace preservation gives $\Tr N_0Z=\Tr M_0Z=0$.
Since $n_0\ne m_0$ and both spectra sum to one, these two equations
imply $Z_{00}=Z_{11}=0$.

For $i=0,1$, define
$K_i:=|i\rangle\langle0|+Z^\dagger|i\rangle\langle1|$ and set
\begin{align}
\widetilde\N(\eta)
&=\sum_{i=0}^1 n_i|i\rangle\langle i|_C\otimes K_i\eta K_i^\dagger
+\langle1|\eta|1\rangle\,|2\rangle\langle2|_C\otimes F,\notag\\
\widetilde\M(\eta)
&=\sum_{i=0}^1 m_i|i\rangle\langle i|_C\otimes K_i\eta K_i^\dagger
+\langle1|\eta|1\rangle\,|2\rangle\langle2|_C\otimes G.
\end{align}
By alignment, summing the branches recovers $\N(\eta)$ and
$\M(\eta)$, respectively. Each branch is completely positive,
and the sums preserve trace, so $\widetilde\N,\widetilde\M$ are channels.
The inward perturbation estimate in Section~\ref{app:inward},
together with one-copy optimality, gives
\begin{align}\label{eq:qubit-inward-bound}
\Tr Z^\dagger\mathfrak B(N_0,M_0)Z+D(F\|G)
&\le D(\N\|\M).
\end{align}
We can therefore bound the relative-entropy increment as follows:
\begin{align}
&D(\widetilde\N(\rho)\|\widetilde\M(\sigma))
-D(\widetilde\N(\rho)\|\widetilde\N(\sigma))\notag\\
&=\sum_{i=0}^1 n_i\ln\frac{n_i}{m_i}
\bigl(r_0+r_1\langle i|ZZ^\dagger|i\rangle\bigr)
+r_1\Tr F(\ln F-\ln G)\notag\\
&=r_0D(\N\|\M)
+r_1\bigl[\Tr Z^\dagger\mathfrak B(N_0,M_0)Z+D(F\|G)\bigr]
\le D(\N\|\M).
\end{align}
The first equality uses the direct-sum formula and
$K_i^\dagger K_i=\diag(1,\langle i|ZZ^\dagger|i\rangle)$;
the trace corrections cancel because both flagged channels preserve trace.
The second equality uses $D(N_0\|M_0)=D(\N\|\M)$ and
$\mathfrak B(N_0,M_0)=N_0(\ln N_0-\ln M_0)-N_0+M_0$.
Here the identities $\Tr F=1-\Tr Z^\dagger N_0Z$ and
$\Tr G=1-\Tr Z^\dagger M_0Z$ cancel the remaining trace corrections.
The inequality follows from \eqref{eq:qubit-inward-bound} and
$r_0+r_1=1$.

\emph{Chain rule and iteration.}
In both cases, the preceding bound yields
\begin{align}
D\bigl((\id_R\otimes\N)(\rho)\big\|
(\id_R\otimes\M)(\sigma)\bigr)
&\le D\bigl((\id_R\otimes\widetilde\N)(\rho)\big\|
(\id_R\otimes\widetilde\M)(\sigma)\bigr)\notag\\
&\le D\bigl((\id_R\otimes\widetilde\N)(\rho)\big\|
(\id_R\otimes\widetilde\N)(\sigma)\bigr)+D(\N\|\M)\notag\\
&\le D(\rho\|\sigma)+D(\N\|\M).
\end{align}
The first inequality discards the branch labels, the second is the
bound proved in each case, and the last is data processing for
$\id_R\otimes\widetilde\N$.
Apply this chain rule successively to the $m$ channel uses, as in
\eqref{eq:sequential-upper}. Starting from the same input under
both hypotheses gives
$\frac1mD(\N^{\otimes m}\|\M^{\otimes m})\le D(\N\|\M)$.
Product inputs give the reverse inequality, so
$D^{\mathrm{reg}}(\N\|\M)=D(\N\|\M)$.
The converse follows from
$2D(\N\|\M)\le D(\N^{\otimes2}\|\M^{\otimes2})\le2D^{\mathrm{reg}}(\N\|\M)$.
\end{proof}

\begin{boxremark}[Failure of universal finite-copy criteria]
\label{rem:finite-copy-criterion}
The two-copy criterion of Theorem~\ref{thm:qubit} cannot be
extended to arbitrary channel dimensions. Indeed, for every $L\ge2$, there exist
channels $\N,\M:A\to B$ with $|A|=3$, $|B|=2$, and
$J_{\N},J_{\M}>0$ such that
\begin{align}
D(\N^{\otimes m}\|\M^{\otimes m})
&=mD(\N\|\M),\qquad 1\le m\le L,\\
D(\N^{\otimes k}\|\M^{\otimes k})
&>kD(\N\|\M)\qquad\text{for some }k>L.
\end{align}

To see this, choose qubit channels $\mathcal E,\mathcal F$ with
positive-definite Choi matrices and
$D^{\mathrm{reg}}(\mathcal E\|\mathcal F)>
D(\mathcal E\|\mathcal F)$; their existence follows from
Theorem~\ref{thm:generic-superadditivity}. Fix $r$ such that
\begin{align}
\max_{1\le j\le L}\frac1j
D(\mathcal E^{\otimes j}\|\mathcal F^{\otimes j})
<r<D^{\mathrm{reg}}(\mathcal E\|\mathcal F).
\end{align}
Such an $r$ exists by Theorem~\ref{thm:universal}, since the finitely
many rates in the maximum are all strictly below
$D^{\mathrm{reg}}(\mathcal E\|\mathcal F)$.
Choose full-rank states $\rho,\sigma$ on $B$ with
$D(\rho\|\sigma)=r$. Identifying
$A\simeq\mathbb C\oplus\mathbb C^2$, define
\begin{align}
\N\!\begin{pmatrix}x&*\\ *&X\end{pmatrix}
=x\rho+\mathcal E(X),\qquad
\M\!\begin{pmatrix}x&*\\ *&X\end{pmatrix}
=x\sigma+\mathcal F(X).
\end{align}
These are channels with positive-definite Choi matrices.
Conditioning on the input block labels and using data processing
and product additivity gives the exact identity
\begin{align}
D(\N^{\otimes n}\|\M^{\otimes n})
&=\max_{0\le j\le n}\bigl\{
D(\mathcal E^{\otimes j}\|\mathcal F^{\otimes j})
+(n-j)r\bigr\},
\end{align}
where the $j=0$ term is $nr$.
For $n\le L$, this maximum equals $nr$, while
for some $k>L$ it strictly exceeds $kr$.
\end{boxremark}

\section{Generic interiority and strict superadditivity}\label{sec:generic}

In this section, we show that the interiority assumption of Theorem~\ref{thm:modular} is generically satisfied (Theorem~\ref{thm:generic-interior}) and that
channel relative entropy is generically strictly superadditive (Theorem~\ref{thm:generic-superadditivity}). Both results are established by showing that any exceptions
are confined to a proper real algebraic subset of the affine space
of trace-preserving Choi pairs. In affine coordinates, this subset
lies in the zero set of a nonzero polynomial and therefore has
Lebesgue measure zero~\cite{Mityagin20}.

Fix systems $A,B$ of dimensions $d_A,d_B\ge2$.
The real affine space of Hermitian Choi pairs satisfying
the trace-preserving constraints is
\begin{align}\label{eq:choi-affine-space}
\mathfrak J_{d_A,d_B}
:=\bigl\{(J_1,J_2):J_j=J_j^\dagger,\
\Tr_B J_j=I_{d_A}\ (j=1,2)\bigr\}.
\end{align}
It has real dimension $2d_A^2(d_B^2-1)$.
The channel Choi pairs form the full-dimensional
convex subset
\begin{align}\label{eq:channel-choi-pairs}
\mathfrak J_{d_A,d_B}^{+}
:=\{(J_1,J_2)\in\mathfrak J_{d_A,d_B}:J_1,J_2\ge0\}.
\end{align}
For $J_{\M}>0$, define the input matrices $K_{\alpha\beta}$ by
\begin{align}\label{eq:input-blocks}
J_{\M}^{-1}J_{\N}=
\sum_{\alpha,\beta=1}^{d_B}
K_{\alpha\beta}\otimes|\alpha\rangle\langle\beta|.
\end{align}
Define the real-linear map on Hermitian input matrices by
\begin{align}\label{eq:linear-obstruction}
\mathcal L(H):=\Tr_B\!\left[
J_{\M}^{-1}J_{\N}(H\otimes I_{d_B})J_{\M}\right].
\end{align}
The operator $\mathcal L(H)$ need not be Hermitian.

\begin{boxassumption}[Generic conditions]\label{ass:generic}
For channel Choi pairs, consider the following conditions:
\begin{enumerate}[label=\textnormal{(G\arabic*)},ref=(G\arabic*)]
\item\label{ass:G1} $J_{\N},J_{\M}>0$.
\item\label{ass:G2} There exists $L\ge1$ such that
products of exactly $L$ matrices from $\{K_{\alpha\beta}\}$
span $\mathcal B(\mathbb C^{d_A})$:
\begin{align}\label{eq:generic-word-span}
\operatorname{span}_{\mathbb C}\!\left\{
K_{\alpha_1\beta_1}\cdots K_{\alpha_L\beta_L}:
1\le\alpha_j,\beta_j\le d_B\right\}=\mathcal B(\mathbb C^{d_A}).
\end{align}
\item\label{ass:G3} The only Hermitian matrix $H$ satisfying
$\mathcal L(H)=H$ is $H=0$.
\end{enumerate}
The last two conditions
are considered only where $J_{\M}>0$. Moreover, denote $\mathcal G_1,\mathcal G_2,\mathcal G_3\subseteq\mathfrak J_{d_A,d_B}^{+}$
for the sets of channel Choi pairs satisfying \ref{ass:G1}, \ref{ass:G2},
and \ref{ass:G3}, respectively. 
\end{boxassumption}

Condition \ref{ass:G2} ensures
that the input blocks of
$(J_{\M}^{-1}J_{\N})^{\otimes m}$ have no nontrivial
common invariant subspace at any blocklength $m$.
Conversely, irreducibility at every blocklength is
equivalent to \ref{ass:G2} by quantum Perron--Frobenius
theory and the quantum Wielandt theorem
\cite{Sanz2010Wielandt,JaquesRahaman2018}.
Condition \ref{ass:G3} is motivated by the modular-flow
criterion in Theorem~\ref{thm:modular}: analytic
continuation to $t=i$ and the Choi representation
yield $\mathcal L(\tau)=\tau$ under two-copy equality.

\subsection{Generic interiority and uniqueness}

We first show that \ref{ass:G1}--\ref{ass:G2} ensure a unique
full-rank optimizer at every blocklength. The key is that the support
of an optimal input marginal is invariant under the input blocks
$K_{\alpha\beta}$ of $J_{\M}^{-1}J_{\N}$. Condition \ref{ass:G2}
then rules out proper invariant subspaces, also at tensor powers.
Full rank and concavity yield uniqueness by analyticity.
Finally, we show that these conditions hold generically by
constructing a Choi pair whose input-block products span
$\mathcal B(\mathbb C^{d_A})$.

\begin{boxlemma}[Invariant optimal supports]
\label{lem:optimal-support}
For a channel pair satisfying \ref{ass:G1}, the support of
every one-copy optimal reference marginal $\tau$ is invariant under
each $K_{\alpha\beta}$.
\end{boxlemma}

\begin{proof}
Let $\tau$ be an optimal marginal and $P$ the orthogonal projection onto $\supp\tau$.
If $P=I$, the claim is trivial. Otherwise, put $Q=I-P$
and write, relative to $(\ran P\oplus\ran Q)\otimes\mathbb C^{d_B}$,
\begin{align}
J_{\N}=\begin{pmatrix}N_0&X\\X^\dagger&N_1\end{pmatrix},
\qquad
J_{\M}=\begin{pmatrix}M_0&Y\\Y^\dagger&M_1\end{pmatrix}.
\end{align}
By Lemma~\ref{lem:general-block-alignment}, $X=N_0Z$ and $Y=M_0Z$
with $Z=M_0^{-1}Y$. Hence
\begin{align}
J_{\M}\begin{pmatrix}M_0^{-1}N_0\\0\end{pmatrix}
=\begin{pmatrix}N_0\\X^\dagger\end{pmatrix}
=J_{\N}\begin{pmatrix}I\\0\end{pmatrix}.
\end{align}
Thus the lower-left block of $J_{\M}^{-1}J_{\N}$ vanishes:
\begin{align}
(Q\otimes I_{d_B})J_{\M}^{-1}J_{\N}(P\otimes I_{d_B})=0,
\end{align}
equivalently $QK_{\alpha\beta}P=0$ for all $\alpha,\beta$.
\end{proof}

\begin{boxtheorem}[Generic interiority and uniqueness]
\label{thm:generic-interior}
For $d_A,d_B\ge2$, let $\mathfrak J_{d_A,d_B}$,
$\mathfrak J_{d_A,d_B}^{+}$, and $\mathcal G_i$
be as in \eqref{eq:choi-affine-space},
\eqref{eq:channel-choi-pairs}, and
Assumption~\ref{ass:generic}, respectively. Then the following hold:
\begin{enumerate}[label=\textnormal{(\roman*)},ref=(\roman*)]
\item\label{thm:generic-interior-algebraic}
The set $\mathfrak J_{d_A,d_B}^{+}\setminus
(\mathcal G_1\cap\mathcal G_2)$
is contained in a proper real algebraic subset of
$\mathfrak J_{d_A,d_B}$.

\item\label{thm:generic-interior-unique}
Every channel pair in $\mathcal G_1\cap\mathcal G_2$
has a unique full-rank maximizer for \eqref{eq:input-optimization},
for every $m\ge1$.
\end{enumerate}
\end{boxtheorem}

In part~\ref{thm:generic-interior-algebraic},
we actually show that all exceptional pairs lie in the zero
set of a nonzero real polynomial on $\mathfrak J_{d_A,d_B}$,
which has Lebesgue measure zero.

\begin{proof}
\textbf{Part \ref{thm:generic-interior-algebraic}: algebraic genericity.}
First, failure of \ref{ass:G1} implies
$\det J_{\N}\det J_{\M}=0$.
This polynomial is nonzero on $\mathfrak J_{d_A,d_B}$,
as witnessed by $J_{\N}=J_{\M}=I_{d_A d_B}/d_B$.
Thus the exceptional set for \ref{ass:G1}
is contained in a proper real algebraic subset.

Next, we construct a witness for \ref{ass:G2}.
Let
\begin{align}
H_0:=\operatorname{diag}(1,\ldots,d_A),\qquad
H_1:=(1,\ldots,1)^{\mathsf T}(1,\ldots,1).
\end{align}
In output-block order, choose
\begin{align}\label{eq:g2-witness}
J_{\M}=\frac{I_{d_A d_B}}{d_B},\qquad
J_{\N}=\frac1{d_B}\left[
\begin{pmatrix}
I_{d_A}+\varepsilon H_1&\varepsilon H_0\\
\varepsilon H_0&I_{d_A}-\varepsilon H_1
\end{pmatrix}
\oplus I_{d_A(d_B-2)}
\right],
\end{align}
where the last summand is omitted if $d_B=2$.
For sufficiently small $\varepsilon>0$, these
are positive definite Choi matrices satisfying
$\Tr_B J_{\N}=\Tr_B J_{\M}=I_{d_A}$.
Since $J_{\M}^{-1}J_{\N}=d_B J_{\N}$,
the input blocks satisfy
\begin{align}
K_{11}&=I_{d_A}+\varepsilon H_1,\qquad
K_{22}=I_{d_A}-\varepsilon H_1,\qquad
K_{12}=\varepsilon H_0.
\end{align}
In particular,
\begin{align}
I_{d_A}=\frac{K_{11}+K_{22}}2,\qquad
H_0=\frac{K_{12}}{\varepsilon},\qquad
H_1=\frac{K_{11}-K_{22}}{2\varepsilon}.
\end{align}
Since $H_0$ has distinct eigenvalues,
Lagrange interpolation gives polynomials
$p_i$ of degree at most $d_A-1$ such that
$p_i(H_0)=|i\rangle\langle i|$.
As every entry of $H_1$ equals $1$,
\begin{align}
p_i(H_0)H_1p_j(H_0)=|i\rangle\langle j|.
\end{align}
Thus products of at most $2d_A-1$ matrices
from $\{I_{d_A},H_0,H_1\}$ span $\mathcal B(\mathbb C^{d_A})$.
Each factor is a linear combination of the
$K_{\alpha\beta}$. Padding shorter products
with copies of $I_{d_A}$ and expanding shows
that products of exactly $2d_A-1$ input blocks
span $\mathcal B(\mathbb C^{d_A})$. Hence the witness
satisfies \ref{ass:G2}.

Finally, we show that failure of \ref{ass:G2}
is algebraically exceptional.
Fix $L=2d_A-1$, and let $\mathsf W_L$ be the
$d_A^2\times d_B^{2L}$ matrix whose columns are
the vectorized length-$L$ products of the
$K_{\alpha\beta}$. The spanning property
is equivalent to $\operatorname{rank}\mathsf W_L=d_A^2$.
We have
\begin{align}
J_{\M}^{-1}J_{\N}
=\frac{\operatorname{adj}(J_{\M})J_{\N}}
       {\det J_{\M}}.
\end{align}
Here $\operatorname{adj}(A)$ denotes the adjugate matrix,
the transpose of the cofactor matrix of $A$.
Its entries are polynomials in the entries of $A$.
Thus each input block has polynomial numerators
and common denominator $\det J_{\M}$.
Consequently, $(\det J_{\M})^L\mathsf W_L$
has polynomial entries in the Choi coordinates
and extends to all of $\mathfrak J_{d_A,d_B}$.
On $\det J_{\M}\ne0$, failure of the
length-$L$ spanning property is equivalent
to the vanishing of all its maximal minors.

By \eqref{eq:g2-witness}, this polynomial
matrix has full rank at the witness.
Hence at least one maximal minor is a
nonzero polynomial on $\mathfrak J_{d_A,d_B}$.
Its squared modulus is a nonzero real
polynomial whose zero set contains the
failure of \ref{ass:G2} whenever
$\det J_{\M}\ne0$.
Combining this with the exceptional set
for \ref{ass:G1} proves
\ref{thm:generic-interior-algebraic}.

\medskip
\textbf{Part \ref{thm:generic-interior-unique}: full-rank uniqueness.}
Fix $m\ge1$ and an optimal marginal $\tau_m$.
Let $P$ be the orthogonal projection onto $\supp\tau_m$ and set $Q=I-P$.
Applying Lemma~\ref{lem:optimal-support} to
$(\N^{\otimes m},\M^{\otimes m})$ gives
\begin{align}
(Q\otimes I_{d_B^m})
(J_{\M}^{-1}J_{\N})^{\otimes m}
(P\otimes I_{d_B^m})=0,
\end{align}
where input and output tensor factors are canonically reordered.
Thus $\ran P$ is invariant under every input block
$K_{\alpha_1\beta_1}\otimes\cdots\otimes
K_{\alpha_m\beta_m}$.
By \ref{ass:G2}, products of some common length $L$
of these blocks span
$\mathcal B(\mathbb C^{d_A})^{\otimes m}=\mathcal B(\mathbb C^{d_A^m})$.
Hence $\ran P$ is invariant under the full matrix algebra,
forcing $P=I$. Every maximizer is therefore full rank.

Suppose distinct states $\tau_0,\tau_1$ are maximizers,
and set $\tau_s=\tau_0+s(\tau_1-\tau_0)$.
Since $\tau_1-\tau_0$ is nonzero, Hermitian, and traceless,
it has a negative eigenvalue, so $\tau_s$ is not positive semidefinite
for sufficiently large $s$.
Thus $\mathcal I=\{s\in\mathbb R:\tau_s>0\}$ is an open interval
containing $[0,1]$, with a finite right endpoint $s_*>1$
at which $\tau_{s_*}$ is singular.

By \ref{ass:G1}, both output matrices defining
$D(\N^{\otimes m},\M^{\otimes m};\tau)$ are positive definite for $\tau>0$.
Since matrix square roots and logarithms are real analytic
on positive definite matrices,
$s\mapsto D(\N^{\otimes m},\M^{\otimes m};\tau_s)$ is real analytic on $\mathcal I$.
Concavity makes it constant on $[0,1]$, so the identity
theorem extends this equality to all $s\in\mathcal I$.
Moreover, \ref{ass:G1} gives
$J_{\N}^{\otimes m}\le\lambda J_{\M}^{\otimes m}$
for some finite $\lambda>0$, ensuring continuity of the
objective also at singular states.
Hence $\tau_{s_*}$ is a maximizer, contradicting the
full-rank property. The maximizer is therefore unique.
\end{proof}

\subsection{Generic strict superadditivity}

We next explain the role of \ref{ass:G3}. By the modular
criterion, two-copy equality forces a full-rank one-copy
optimizer to be a fixed point of $\mathcal L$. Thus
\ref{ass:G3} excludes equality at two copies. We show
that this condition is algebraically generic using an
explicit Choi-pair witness. Uniqueness from the preceding
subsection then extends strict superadditivity to every
pair of blocklengths, and finite nonattainment follows.

\begin{boxlemma}[Linear additivity obstruction]
\label{lem:linear-obstruction}
For a channel pair satisfying \ref{ass:G1}, let $\tau>0$
maximize the one-copy objective in \eqref{eq:input-optimization}.
If $D(\N^{\otimes 2}\|\M^{\otimes 2})=2D(\N\|\M)$,
then $\mathcal L(\tau)=\tau$.
\end{boxlemma}

\begin{proof}
Let
\begin{align}
\rho=(\sqrt\tau\otimes I_{d_B})J_{\N}(\sqrt\tau\otimes I_{d_B}),
\qquad
\sigma=(\sqrt\tau\otimes I_{d_B})J_{\M}(\sqrt\tau\otimes I_{d_B}).
\end{align}
By Theorem~\ref{thm:modular},
$\Tr_B(\sigma^{it}\rho\sigma^{-it})=\tau$
for every real $t$. Since $\sigma>0$,
the matrix powers $\sigma^{iz}=e^{iz\ln\sigma}$
and $\sigma^{-iz}=e^{-iz\ln\sigma}$ are entire
in $z\in\mathbb C$. Matrix multiplication and
partial trace preserve holomorphy, so
$z\mapsto\Tr_B(\sigma^{iz}\rho\sigma^{-iz})$
is entire. By the identity theorem, the
equality also holds at $z=i$. Hence
\begin{align}
\tau=\Tr_B(\sigma^{-1}\rho\sigma)
=\tau^{-1/2}\mathcal L(\tau)\tau^{1/2},
\end{align}
which gives $\mathcal L(\tau)=\tau$.
\end{proof}

\begin{boxtheorem}[Generic strict superadditivity and finite nonattainment]
\label{thm:generic-superadditivity}
For $d_A,d_B\ge2$, let $\mathfrak J_{d_A,d_B}$,
$\mathfrak J_{d_A,d_B}^{+}$, and $\mathcal G_i$
be as in \eqref{eq:choi-affine-space},
\eqref{eq:channel-choi-pairs}, and
Assumption~\ref{ass:generic}, respectively. Then the following hold:
\begin{enumerate}[label=\textnormal{(\roman*)},ref=(\roman*)]
\item\label{thm:generic-superadditivity-algebraic}
The set $\mathfrak J_{d_A,d_B}^{+}\setminus
(\mathcal G_1\cap\mathcal G_2\cap\mathcal G_3)$
is contained in a proper real algebraic subset of
$\mathfrak J_{d_A,d_B}$.

\item\label{thm:generic-superadditivity-strict}
Every channel pair in
$\mathcal G_1\cap\mathcal G_2\cap\mathcal G_3$ satisfies
\begin{align}
D(\N^{\otimes(m+n)}\|\M^{\otimes(m+n)})
>D(\N^{\otimes m}\|\M^{\otimes m})
+D(\N^{\otimes n}\|\M^{\otimes n}),\qquad m,n\ge1.
\end{align}

\item\label{thm:generic-superadditivity-nonattainment}
As a consequence of \ref{thm:generic-superadditivity-strict},
every channel pair in
$\mathcal G_1\cap\mathcal G_2\cap\mathcal G_3$ satisfies
\begin{align}\label{eq:generic-nonattainment}
D^{\mathrm{reg}}(\N\|\M) > \frac1mD(\N^{\otimes m}\|\M^{\otimes m})
\qquad m\ge1.
\end{align}
\end{enumerate}
\end{boxtheorem}

In part~\ref{thm:generic-superadditivity-algebraic},
we actually show that all exceptional pairs lie in the zero set of a nonzero real polynomial on $\mathfrak J_{d_A,d_B}$,
which has Lebesgue measure zero.

\begin{proof}
\textbf{Part \ref{thm:generic-superadditivity-algebraic}: algebraic genericity.}
By Theorem~\ref{thm:generic-interior},
part~\ref{thm:generic-interior-algebraic},
it remains to establish the algebraic genericity
of \ref{ass:G3}.

First, we construct a witness at which
$\mathcal L-\id$ is injective on Hermitian matrices.
The witness matrices need not be positive semidefinite, since
$\mathfrak J_{d_A,d_B}^{+}$ is full-dimensional.
Let
\begin{align}
H_0:=\operatorname{diag}(1,\ldots,d_A),\qquad
H_1:=(1,\ldots,1)^{\mathsf T}(1,\ldots,1),\qquad
K:=\bigl((1-\mathrm i)I_{d_A}+(1+\mathrm i)H_1\bigr)^{-1}.
\end{align}
The matrix defining $K$ is invertible, since its
Hermitian part is $I_{d_A}+H_1>0$.
Writing the Choi matrices in $d_A\times d_A$ blocks, choose
\begin{align}
J_{\M}&=
\begin{pmatrix}(d_B-1)I_{d_A}&H_0\\H_0&0\end{pmatrix}
\oplus(-I_{d_A(d_B-2)}),\qquad
J_{\N}=J_{\M}+
\left[
\begin{pmatrix}0&K^\dagger H_0\\H_0K&0\end{pmatrix}
\oplus0_{d_A(d_B-2)}
\right],
\end{align}
omitting the last summands if $d_B=2$.
Both matrices are Hermitian with partial trace $I_{d_A}$,
and $J_{\M}$ is invertible since $H_0$ is invertible.
Direct block multiplication gives
\begin{align}
\mathcal L(H)-H
&=\Tr_B\!\left[
J_{\M}^{-1}(J_{\N}-J_{\M})
(H\otimes I_{d_B})J_{\M}\right]
=(d_B-1)\bigl(KH-H_0^{-1}KHH_0\bigr).
\end{align}
If $H=H^\dagger$ and $\mathcal L(H)=H$,
then $H_0KH=KHH_0$.
Since $H_0$ has distinct diagonal entries,
$KH=\operatorname{diag}(z_1,\ldots,z_{d_A})$.
Thus
\begin{align}
H=K^{-1}\operatorname{diag}(z_1,\ldots,z_{d_A}),
\qquad
H_{jj}=2z_j,\quad
H_{jk}=(1+\mathrm i)z_k\quad(j\ne k).
\end{align}
Since $d_A\ge2$, Hermiticity gives $z_j\in\mathbb R$
and $(1+\mathrm i)z_k=(1-\mathrm i)z_j$
for $j\ne k$. Comparing real and imaginary parts
yields $z_k=z_j=-z_j$, hence all $z_j=0$.
Thus $H=0$, proving injectivity.

Next, we show that failure of \ref{ass:G3}
is algebraically exceptional. For $\det J_{\M}\ne0$,
\begin{align}
(\det J_{\M})(\mathcal L-\id)(H)
=\Tr_B\!\left[
\operatorname{adj}(J_{\M})J_{\N}
(H\otimes I_{d_B})J_{\M}\right]
-(\det J_{\M})H.
\end{align}
As a real-linear map from $\operatorname{Herm}_{d_A}$
to $\mathcal B(\mathbb C^{d_A})$, the right-hand side is represented
by a $2d_A^2\times d_A^2$ real matrix with entries
polynomial in the Choi coordinates.
At the witness, $\mathcal L-\id$ is injective
and $\det J_{\M}\ne0$, so this matrix has rank $d_A^2$.
Thus some maximal minor is a nonzero real
polynomial $p$ on $\mathfrak J_{d_A,d_B}$.
For any pair with $\det J_{\M}\ne0$ failing
\ref{ass:G3}, the matrix has rank less than
$d_A^2$, so $p=0$.
Hence all failures of \ref{ass:G3} are contained
in the zero set of the nonzero polynomial
$(\det J_{\M})p$.
This proves \ref{thm:generic-superadditivity-algebraic}.

\medskip
\textbf{Part \ref{thm:generic-superadditivity-strict}: strict superadditivity.}
Fix a pair in $\mathcal G_1\cap\mathcal G_2\cap\mathcal G_3$.
By Theorem~\ref{thm:generic-interior},
part~\ref{thm:generic-interior-unique},
every blocklength $m\ge1$ has a unique full-rank
maximizing marginal $\tau_m$.
If $D(\N^{\otimes2}\|\M^{\otimes2})=2D(\N\|\M)$,
Lemma~\ref{lem:linear-obstruction} gives
$\mathcal L(\tau_1)=\tau_1$, contradicting
\ref{ass:G3}. Hence
$D(\N^{\otimes2}\|\M^{\otimes2})>2D(\N\|\M)$.

Suppose equality in superadditivity holds
for some $m,n\ge1$:
\begin{align}
D(\N^{\otimes(m+n)}\|\M^{\otimes(m+n)})
=D(\N^{\otimes m}\|\M^{\otimes m})
+D(\N^{\otimes n}\|\M^{\otimes n}).
\end{align}
Then $\tau_m\otimes\tau_n$ is an optimal
$(m+n)$-copy marginal. Since all channel uses
are identical, permuting the input sites
only permutes the corresponding output sites,
leaving the objective unchanged.
By uniqueness, $\tau_m\otimes\tau_n$ is therefore
invariant under every permutation of its $m+n$ sites.

Let $\omega$ be the reduced state of any single
site, which is the same for all sites by symmetry.
For any site $i$ in the first block, exchange it
with a site in the second block and trace out
the second block. Since the blocks are initially
in a product state, permutation invariance gives
$\tau_m=\omega_i\otimes\Tr_i\tau_m$,
where $\omega_i$ acts on site $i$.
Thus each site is independent of the remaining
sites. Applying this to both blocks yields
$\tau_m=\omega^{\otimes m}$,
$\tau_n=\omega^{\otimes n}$.
By additivity on product inputs,
\begin{align}
D(\N^{\otimes(m+n)}\|\M^{\otimes(m+n)})
=(m+n)D(\N,\M;\omega)
\le(m+n)D(\N\|\M).
\end{align}
But superadditivity and the strict two-copy
inequality give
\begin{align}
D(\N^{\otimes(m+n)}\|\M^{\otimes(m+n)})
&\ge D(\N^{\otimes2}\|\M^{\otimes2})
+(m+n-2)D(\N\|\M)
>(m+n)D(\N\|\M),
\end{align}
a contradiction.

\medskip
\textbf{Part \ref{thm:generic-superadditivity-nonattainment}: finite nonattainment.}
By part~\ref{thm:generic-superadditivity-strict},
$D(\N^{\otimes 2m}\|\M^{\otimes 2m})
>2D(\N^{\otimes m}\|\M^{\otimes m})$ for every $m\ge1$.
Since $D^{\mathrm{reg}}(\N\|\M)\ge
D(\N^{\otimes 2m}\|\M^{\otimes 2m})/(2m)$,
the claim follows.
\end{proof}

\section{Efficient approximation of the regularized divergence}
\label{sec:regularized-computation}

In this section, we establish quantitative approximation guarantees
for the regularized channel relative entropy.
We show that $\widetilde O(\varepsilon^{-3/2})$ channel copies suffice
to approximate $D^{\mathrm{reg}}(\N\|\M)$ from below to additive
accuracy $\varepsilon$ (Theorem~\ref{thm:physical-block}).
This blocklength bound follows by combining the variational
characterization of~\cite{SS26} with spectral localization.
We also obtain a certified upper approximation with the same accuracy
at level $\widetilde O(\varepsilon^{-2})$ of the sharp R\'enyi hierarchy
(Theorem~\ref{thm:upper-hierarchy}).
Quantitative R\'enyi continuity and the symmetry reduction
of~\cite{FF21,FST22} allow this upper bound to be evaluated
to the required accuracy in polynomial time.

\subsection{Quantitative R\'enyi continuity}
For $\alpha>1$, the sandwiched R\'enyi divergence of states
\cite{MLDS13,WWY14} is
\begin{align}\label{eq:s7-sandwiched-renyi}
\widetilde D_\alpha(\rho\|\sigma)
&:=\frac1{\alpha-1}\ln\Tr\left(
\sigma^{\frac{1-\alpha}{2\alpha}}\rho
\sigma^{\frac{1-\alpha}{2\alpha}}\right)^\alpha,
\end{align}
if $\supp\rho\subseteq\supp\sigma$ and 
$+\infty$ otherwise.
The negative powers of $\sigma$ are evaluated on $\supp\sigma$
and extended by zero on its orthogonal complement.
The corresponding channel divergence and its regularization are
defined by replacing $D$ with $\widetilde D_\alpha$ in
\eqref{eq:channel-relative-entropy} and
\eqref{eq:channel-relative-entropy-regularized}.
Continuity of the regularized divergence at $\alpha=1$
was recently established in \cite{GJL26,SS26,WY26,ZW26}.

Using the variational characterization of
\cite[Lemmas~3.1--3.2]{SS26}, we obtain a quantitative
Rényi continuity bound with an explicit choice of order
under a lower bound on $J_{\M}$.

\begin{boxlemma}[Quantitative R\'enyi continuity]
\label{lem:renyi-continuity}
Let $\N,\M:A\to B$ be quantum channels with
$|A|=d_A$ and $|B|=d_B\ge2$, and suppose
$J_{\M}\ge\kappa I_{d_A d_B}/d_B$ for some $0<\kappa\le1$.
Then, for every $0<\varepsilon\le1$ and
$0<s\le\varepsilon/
(128[1+\ln(d_B^2/(\kappa\varepsilon))]^2)$,
\begin{align}\label{eq:renyi-continuity}
0&\le
\widetilde D_{1/(1-s)}^{\mathrm{reg}}(\N\|\M)
-D^{\mathrm{reg}}(\N\|\M)
\le\frac{\varepsilon}{2}.
\end{align}
\end{boxlemma}

\begin{proof}
Write $\N_R=\id_R\otimes\N$ and $\M_R=\id_R\otimes\M$.
By \cite[Lemmas~3.1--3.2]{SS26}, for $0<s<1$,
\begin{align}
C_s&:=\sup_{R,T>0}\frac1s\ln\left\|
[\M_R^*(T)]^{-s/2}\N_R^*(T^s)[\M_R^*(T)]^{-s/2}
\right\|_\infty
=\widetilde D_{1/(1-s)}^{\mathrm{reg}}(\N\|\M),\label{eq:s7-Cs}\\
C_0&:=\sup_{R,T>0}\lambda_{\max}\bigl(\N_R^*(\ln T)-\ln\M_R^*(T)\bigr)
=D^{\mathrm{reg}}(\N\|\M),\label{eq:s7-C0}
\end{align}
where $R$ is any finite reference system and $T>0$ acts on
$R\otimes B$.
Here $^*$ denotes the Hilbert--Schmidt adjoint, characterized by
$\Tr[X\N(\rho)]=\Tr[\N^*(X)\rho]$ for every input state $\rho$
and output observable $X$.
In particular, $\N_R^*=\id_R\otimes\N^*$ and
$\M_R^*=\id_R\otimes\M^*$ map operators on $R\otimes B$
to operators on $R\otimes A$.
The adjoint of a quantum channel is completely positive and unital;
for example, $\N^*(I_B)=I_A$.

Since $C_s$ is nondecreasing in $s$, it suffices
to consider the largest admissible value. Set
\begin{align}\label{eq:s7-parameters}
h:=1+\ln\frac{d_B^2}{\kappa\varepsilon},\qquad
s:=\frac{\varepsilon}{128h^2},\qquad
w:=\ln\left(1+\frac{d_B^2}{\kappa s}\right).
\end{align}

Fix $R,T>0$, and set
$U:=T+\kappa s(\Tr_B T/d_B)\otimes I_{d_B}$, $\mathcal R(X):=\Tr(X)I_{d_B}/d_B$.
The Choi bound gives $\M\ge_{\mathrm{CP}}\kappa\mathcal R$,
while pinching on $B$ yields
\begin{align}\label{eq:s7-basic-bounds}
\M_R^*(T)\ge\kappa\frac{\Tr_B T}{d_B}\otimes I_{d_A},
\qquad T\le d_B(\Tr_B T)\otimes I_{d_B}.
\end{align}
Consequently, using unitality of $\M_R^*$,
\begin{align}
\kappa s\frac{\Tr_B T}{d_B}\otimes I_{d_B}
&\le U\le(d_B^2+\kappa s)\frac{\Tr_B T}{d_B}\otimes I_{d_B},
\label{eq:s7-U-bounds}\\
\kappa s\frac{\Tr_B T}{d_B}\otimes I_{d_A}
&\le\M_R^*(U)\le(d_B^2+\kappa s)\frac{\Tr_B T}{d_B}\otimes I_{d_A},
\label{eq:s7-MU-bounds}\\
\M_R^*(T)&\le\M_R^*(U)\le(1+s)\M_R^*(T).
\label{eq:s7-M-comparison}
\end{align}
The last inequality also uses
$\M_R^*(U)=\M_R^*(T)+\kappa s(\Tr_B T/d_B)\otimes I_{d_A}$.
By \eqref{eq:s7-U-bounds}--\eqref{eq:s7-MU-bounds},
positivity and unitality of $\N_R^*$, and
the order-reversing property of inversion,
\begin{align}
\N_R^*(U)
&\le(d_B^2+\kappa s)\frac{\Tr_B T}{d_B}\otimes I_{d_A}
\le e^w\M_R^*(U),\label{eq:s7-positive-bounds}\\
\N_R^*(U^{-1})
&\le\frac1{\kappa s}\left(\frac{\Tr_B T}{d_B}\right)^{-1}\!\otimes I_{d_A}
\le e^w[\M_R^*(U)]^{-1}.\label{eq:s7-inverse-bounds}
\end{align}

Define the operator-valued analytic function
\begin{align}\label{eq:s7-G}
G(z):=[\M_R^*(U)]^{-z/2}\N_R^*(U^z)[\M_R^*(U)]^{-z/2}.
\end{align}
As $\N_R^*$ is unital and completely positive,
$\|G(it)\|_\infty\le1$. 
By the Stinespring representation
$\N_R^*(X)=V^\dagger(X\otimes I_E)V$
and the factorization
$U^{\pm1+it}=U^{\pm1/2}U^{it}U^{\pm1/2}$,
Cauchy--Schwarz gives
\begin{align}\label{eq:s7-boundary}
\|G(\pm1+it)\|_\infty
\le\left\|[\M_R^*(U)]^{\mp1/2}
\N_R^*(U^{\pm1})
[\M_R^*(U)]^{\mp1/2}\right\|_\infty
\le e^w.
\end{align}
The three-lines theorem on the two adjacent strips yields
\begin{align}\label{eq:s7-three-lines}
\|G(z)\|_\infty\le e^{w|\Re z|}
\qquad(|\Re z|\le1).
\end{align}

By \eqref{eq:s7-parameters}, $3\le w\le4h$ and
$sw\le1/2$. Thus the circle $|z|=2/w$ lies in
the strip, where $\|G(z)\|_\infty\le e^2$.
Since
\begin{align}\label{eq:s7-G-derivative}
G(0)=I,\qquad
G'(0)=\N_R^*(\ln U)-\ln\M_R^*(U),
\end{align}
Cauchy's coefficient estimate and the geometric series
bound, using $sw/2\le1/4$, give
\begin{align}\label{eq:s7-Cauchy}
\|G(s)-I-sG'(0)\|_\infty
\le\frac{e^2s^2w^2}{4(1-sw/2)}
\le\frac{e^2}{3}s^2w^2<3s^2w^2.
\end{align}

By \eqref{eq:s7-Cauchy} and the definition of $C_0$,
$G(s)\le(1+sC_0+3s^2w^2)I
\le e^{sC_0+3s^2w^2}I$. By operator monotonicity of
$X^s$ for $0<s<1$ and \eqref{eq:s7-M-comparison},
\begin{align}\label{eq:s7-final-comparison}
\N_R^*(T^s)&\le\N_R^*(U^s)
\le e^{sC_0+3s^2w^2}[\M_R^*(U)]^s
\le e^{sC_0+3s^2w^2}(1+s)^s[\M_R^*(T)]^s.
\end{align}
Conjugating by $[\M_R^*(T)]^{-s/2}$, taking
the operator norm, and using the formula for $C_s$, we obtain
\begin{align}\label{eq:s7-final-error}
0\le C_s-C_0\le\ln(1+s)+3sw^2
\le\frac{\varepsilon}{128h^2}+\frac{3\varepsilon}{8}
<\frac{\varepsilon}{2}.
\end{align}
Monotonicity of $C_s$ extends the bound to all smaller $s$.
\end{proof}

\begin{boxremark}[Continuity at order one]
\label{rem:renyi-continuity-limit}
Lemma~\ref{lem:renyi-continuity} recovers right continuity at
$\alpha=1$ whenever $J_{\M}>0$:
\begin{align}\label{eq:renyi-continuity-limit}
\lim_{\alpha\to1^+}
\widetilde D_\alpha^{\mathrm{reg}}(\N\|\M)
&=D^{\mathrm{reg}}(\N\|\M).
\end{align}
Indeed, fix an admissible $\kappa>0$ and set $s=1-\alpha^{-1}$.
For every $0<\varepsilon\le1$, the hypothesis on $s$ holds for
all $\alpha>1$ sufficiently close to $1$.
Equation~\eqref{eq:renyi-continuity} then bounds the divergence
gap by $\varepsilon/2$, proving the limit.
For fixed $d_B,\kappa$, the convergence is uniform over channel
pairs satisfying $J_{\M}\ge\kappa I_{d_A d_B}/d_B$.
\end{boxremark}

\subsection{Finite-block approximation}
\label{subsec:physical-block}

\begin{boxtheorem}[Finite-block approximation]
\label{thm:physical-block}
Let $\N,\M:A\to B$ be quantum channels with
$|A|=d_A$, $|B|=d_B\ge2$, and
$J_{\M}\ge\kappa I_{d_A d_B}/d_B$ for some $0<\kappa\le1$.
For $0<\varepsilon\le1$, any integer $m$ with
\begin{align}\label{eq:physical-explicit-length}
m&\ge24\sqrt3\,\pi
\left[\frac{2+\ln(1+2d_B^2/(\kappa\varepsilon))}
{\varepsilon}\right]^{3/2}
\end{align}
satisfies
\begin{align}\label{eq:physical-block-error}
0&\le D^{\mathrm{reg}}(\N\|\M)
-\frac1mD(\N^{\otimes m}\|\M^{\otimes m})
<\varepsilon.
\end{align}
In particular, the sufficient blocklength is
$\widetilde O(\varepsilon^{-3/2})$ for fixed $d_B,\kappa$.
\end{boxtheorem}

The proof uses the variational formula \eqref{eq:s7-C0} and
spectral localization, following the confinement approach
of~\cite[Section~3.3]{SS26}.
We first regularize the variational test and bound the double
commutator in the following lemma through the curvature of a
unitary conjugation. This bound is uniform in the reference dimension.
We then localize a tensor-power input to a spectral interval of
width $L$. A cosine window bounds the loss in the per-copy
variational value by $O(L^{-2})$, while the conversion to the channel
divergence costs at most $L/m$.
Balancing these terms gives the stated blocklength bound.
The following lemma provides the localization estimate.

\begin{boxlemma}[Spectral localization]
\label{lem:spectral-localization}
Let $Y,H$ be self-adjoint operators on a finite-dimensional
Hilbert space, let $|\psi\rangle$ be a unit vector, and let $L>0$.
There exist $x\in\mathbb R$ and a unit vector $|\xi\rangle$
in the spectral subspace of $Y$ corresponding to
$[x-L/2,x+L/2]$ such that
\begin{align}\label{eq:spectral-localization}
\langle\xi|H|\xi\rangle
&\ge\langle\psi|H|\psi\rangle
-\frac{\pi^2}{2L^2}\|[Y,[Y,H]]\|_\infty.
\end{align}
\end{boxlemma}

\begin{proof}
Define the normalized cosine window and its translates by
\begin{align}\label{eq:cosine-window}
f_L(u)&:=\sqrt{\frac2L}\cos\frac{\pi u}{L}\,
\mathbf1_{[-L/2,L/2]}(u),
&K_x&:=f_L(Y-xI).
\end{align}
We use the unitary Fourier transform
$\widehat f_L(t)=(2\pi)^{-1/2}\int_{\mathbb R}e^{-itu}f_L(u)\,\mathrm du$.
The window is continuous, with square-integrable weak derivative,
and Plancherel's identity gives
\begin{align}\label{eq:cosine-moments}
\int_{\mathbb R}|\widehat f_L(t)|^2\,\mathrm dt&=1,
&\int_{\mathbb R}t^2|\widehat f_L(t)|^2\,\mathrm dt
&=\int_{\mathbb R}|f_L'(u)|^2\,\mathrm du=\frac{\pi^2}{L^2}.
\end{align}
The spectral theorem and the Fourier autocorrelation identity imply
\begin{align}\label{eq:localization-average}
\int_{\mathbb R}K_x^2\,\mathrm dx&=I,\notag\\
\int_{\mathbb R}K_xHK_x\,\mathrm dx
&=\int_{\mathbb R}|\widehat f_L(t)|^2
 e^{itY}He^{-itY}\,\mathrm dt.
\end{align}
Since $|\widehat f_L(t)|^2$ is even, the linear term in $t$
in the second integral vanishes. The second-order Taylor
remainder and \eqref{eq:cosine-moments} therefore yield
\begin{align}\label{eq:localization-average-error}
\left\|\int_{\mathbb R}K_xHK_x\,\mathrm dx-H\right\|_\infty
&\le\frac12\|[Y,[Y,H]]\|_\infty
\int_{\mathbb R}t^2|\widehat f_L(t)|^2\,\mathrm dt\notag\\
&=\frac{\pi^2}{2L^2}\|[Y,[Y,H]]\|_\infty.
\end{align}
Indeed, the second derivative of $e^{itY}He^{-itY}$ is
$-e^{itY}[Y,[Y,H]]e^{-itY}$, whose operator norm is independent of $t$.

Set $p_x:=\|K_x|\psi\rangle\|^2$.
Equation~\eqref{eq:localization-average} gives
$\int_{\mathbb R}p_x\,\mathrm dx=1$.
Taking the expectation of \eqref{eq:localization-average-error}
in $|\psi\rangle$ shows that, for some $x$ with $p_x>0$,
the unit vector $|\xi\rangle:=p_x^{-1/2}K_x|\psi\rangle$
satisfies \eqref{eq:spectral-localization}.
Its spectral support lies in the stated interval by
\eqref{eq:cosine-window}.
\end{proof}

\begin{proof}[Proof of Theorem~\ref{thm:physical-block}]
Fix $\eta>0$, a finite reference $R$, a positive definite test
$T$ on $R\otimes B$, and a unit vector $|\psi\rangle$ on
$R\otimes A$. We derive a bound uniform in these variational
choices and then take the supremum in \eqref{eq:s7-C0}.

Set
\begin{align}\label{eq:physical-smoothed-test}
S&:=\frac{\Tr_B T}{d_B},
&U&:=T+\eta S\otimes I_B,
&P&:=(\id_R\otimes\M)^*(U),\notag\\
Y&:=\ln P,
&H&:=(\id_R\otimes\N)^*(\ln U)-Y,
&w_\eta&:=\ln\left(1+\frac{d_B^2}{\eta}\right).
\end{align}
Here $S,U,P$ are positive definite.
The estimates \eqref{eq:s7-U-bounds}--\eqref{eq:s7-M-comparison}
in the proof of Lemma~\ref{lem:renyi-continuity} apply with
$\kappa s$ replaced by $\eta$.
In particular, $U\ge T$ and
$P\le(1+\eta/\kappa)(\id_R\otimes\M)^*(T)$.
Operator monotonicity of $\ln$ therefore gives the regularization loss
\begin{align}\label{eq:physical-smoothing-loss}
H&\ge(\id_R\otimes\N)^*(\ln T)
-\ln\bigl[(\id_R\otimes\M)^*(T)\bigr]
-\ln(1+\eta/\kappa)I_{RA}.
\end{align}

\emph{Uniform curvature bound.}
Let $V=I_R\otimes V_{\N}$, where
$V_{\N}:A\to B\otimes E$ is a Stinespring isometry for $\N$,
and define
\begin{align}\label{eq:physical-transfer}
F(z)&:=(U^z\otimes I_E)VP^{-z},\qquad z\in\mathbb C.
\end{align}
Likewise, \eqref{eq:s7-positive-bounds}--\eqref{eq:s7-inverse-bounds}
give
\begin{align}\label{eq:physical-transfer-endpoints}
(\id_R\otimes\N)^*(U)&\le e^{w_\eta}P,
&(\id_R\otimes\N)^*(U^{-1})&\le e^{w_\eta}P^{-1}.
\end{align}
Thus $\|F(\pm1/2+it)\|_\infty\le e^{w_\eta/2}$ and
$\|F(it)\|_\infty=1$ for every $t\in\mathbb R$.
The imaginary powers are unitary, so these bounds hold along
the entire vertical lines. The three-lines theorem yields
\begin{align}\label{eq:physical-transfer-strip}
\|F(z)\|_\infty&\le e^{w_\eta|\Re z|}
\qquad(|\Re z|\le1/2).
\end{align}
Cauchy's estimate on the circle $|z|=(w_\eta+2)^{-1}$ then gives
\begin{align}\label{eq:physical-transfer-derivatives}
\|F^{(j)}(0)\|_\infty&\le j!e(w_\eta+2)^j,
\qquad j=1,2,3.
\end{align}
The connection to spectral localization is the identity
\begin{align}\label{eq:physical-unitary-orbit}
F(it)^\dagger F'(it)&=e^{itY}He^{-itY},\qquad t\in\mathbb R.
\end{align}
It follows directly from $H=V^\dagger(\ln U\otimes I_E)V-Y$.
Differentiating twice with respect to $t$ gives
\begin{align}\label{eq:physical-commutator-bound}
\|[Y,[Y,H]]\|_\infty
&=\left\|\left.\frac{\mathrm d^2}{\mathrm dt^2}
\bigl(F(it)^\dagger F'(it)\bigr)\right|_{t=0}\right\|_\infty\notag\\
&\le\|F'''(0)\|_\infty
+3\|F'(0)\|_\infty\|F''(0)\|_\infty\notag\\
&\le6e(1+e)(w_\eta+2)^3
<64(w_\eta+2)^3.
\end{align}
The first inequality follows from the product rule and $\|F(0)\|_\infty=1$;
the second uses \eqref{eq:physical-transfer-derivatives}.
This bounds the curvature of $e^{itY}He^{-itY}$ uniformly
in the reference system and the variational test.

\emph{Spectral localization.}
On $(R\otimes A)^{\otimes m}$, write
$Y_m:=\sum_{j=1}^mY_j$ and $H_m:=\sum_{j=1}^mH_j$,
where the subscript $j$ indicates the tensor factor on which
an operator acts. Operators on distinct factors commute, so
\begin{align}\label{eq:physical-tensor-commutator}
[Y_m,[Y_m,H_m]]&=\sum_{j=1}^m[Y_j,[Y_j,H_j]],
&\|[Y_m,[Y_m,H_m]]\|_\infty&\le64m(w_\eta+2)^3.
\end{align}
Apply Lemma~\ref{lem:spectral-localization} to $Y_m,H_m$ and
$|\psi\rangle^{\otimes m}$. For every $L>0$, it supplies a
unit vector $|\xi\rangle$, supported in a $Y_m$-spectral interval
of width $L$, such that
\begin{align}\label{eq:physical-localized-input}
\frac1m\langle\xi|H_m|\xi\rangle
&\ge\langle\psi|H|\psi\rangle
-\frac{32\pi^2(w_\eta+2)^3}{L^2}.
\end{align}
The spectral support also gives
\begin{align}\label{eq:physical-spectral-penalty}
0&\le\ln\langle\xi|e^{Y_m}|\xi\rangle
-\langle\xi|Y_m|\xi\rangle\le L.
\end{align}
Indeed, both terms in the difference lie between the endpoints
of that interval, and the lower bound follows from Jensen's inequality.

Let $\rho_\xi$ and $\sigma_\xi$ be the outputs of
$\id_{R^{\otimes m}}\otimes\N^{\otimes m}$ and
$\id_{R^{\otimes m}}\otimes\M^{\otimes m}$, respectively,
on $|\xi\rangle\langle\xi|$.
Measuring the spectral projections of a positive definite test
$Q$ and applying the classical variational inequality gives
$D(\rho_\xi\|\sigma_\xi)\ge
\Tr(\rho_\xi\ln Q)-\ln\Tr(\sigma_\xi Q)$.
With $Q=U^{\otimes m}$, we therefore have
\begin{align}\label{eq:physical-localization-conversion}
D(\N^{\otimes m}\|\M^{\otimes m})
&\ge D(\rho_\xi\|\sigma_\xi)\notag\\
&\ge\langle\xi|(H_m+Y_m)|\xi\rangle
-\ln\langle\xi|e^{Y_m}|\xi\rangle\notag\\
&\ge m\langle\psi|H|\psi\rangle
-\frac{32\pi^2m(w_\eta+2)^3}{L^2}-L.
\end{align}
Here $(\id_{R^{\otimes m}}\otimes\M^{\otimes m})^*
(U^{\otimes m})=P^{\otimes m}=e^{Y_m}$.
The first inequality allows an arbitrary finite reference;
equivalently, Schmidt decomposition compresses the reference of
$|\xi\rangle$ to dimension at most $d_A^m$ without changing
the output relative entropy.

Combining \eqref{eq:physical-localization-conversion} with
\eqref{eq:physical-smoothing-loss} and taking the supremum over
$R,T,|\psi\rangle$ in \eqref{eq:s7-C0} gives
\begin{align}\label{eq:physical-error-balance}
D^{\mathrm{reg}}(\N\|\M)
-\frac1mD(\N^{\otimes m}\|\M^{\otimes m})
&\le\ln\left(1+\frac{\eta}{\kappa}\right)
+\frac{32\pi^2(w_\eta+2)^3}{L^2}+\frac Lm.
\end{align}
All error terms are uniform in these variational choices, so
the supremum need not be attained.
The three terms account for regularization of the test, spectral
localization, and conversion to the channel divergence, respectively.
Balancing the last two terms by choosing
$L=4\pi^{2/3}(w_\eta+2)m^{1/3}$ gives
\begin{align}\label{eq:physical-localization-error}
0&\le D^{\mathrm{reg}}(\N\|\M)
-\frac1mD(\N^{\otimes m}\|\M^{\otimes m})\notag\\
&\le\ln\left(1+\frac{\eta}{\kappa}\right)
+\frac{6\pi^{2/3}}{m^{2/3}}
\left[2+\ln\left(1+\frac{d_B^2}{\eta}\right)\right].
\end{align}
The lower bound follows from the definition of regularization.
Finally, set $\eta=\kappa\varepsilon/2$.
Then $\ln(1+\eta/\kappa)<\varepsilon/2$, while
\eqref{eq:physical-explicit-length} makes the second term in
\eqref{eq:physical-localization-error} at most $\varepsilon/2$, proving
\eqref{eq:physical-block-error}.
\end{proof}

The amortized channel relative entropy is defined by
\begin{align}\label{eq:amortized-divergence}
D^A(\N\|\M)
:=\sup_{\substack{\rho,\sigma\in\mathcal D(R\otimes A)\\
D(\rho\|\sigma)<\infty}}
\left[
D\bigl((\id_R\otimes\N)(\rho)\big\|
(\id_R\otimes\M)(\sigma)\bigr)
-D(\rho\|\sigma)
\right],
\end{align}
where $R$ ranges over all finite-dimensional reference
systems. For the Umegaki relative entropy,
\cite[Corollary~3]{FFRS} established
$D^A(\N\|\M)=D^{\mathrm{reg}}(\N\|\M)$.
Our finite-block estimate also gives a bound on the
reference dimension needed to approximate this value.
For $r\ge1$, define
\begin{align}\label{eq:bounded-amortized}
D^{A,(r)}(\N\|\M)
&:=\sup_{\substack{|R|\le r,\;
\rho,\sigma\in\mathcal D(R\otimes A)\\
D(\rho\|\sigma)<\infty}}
\left[
D\bigl((\id_R\otimes\N)(\rho)\big\|
(\id_R\otimes\M)(\sigma)\bigr)
-D(\rho\|\sigma)
\right].
\end{align}

\begin{boxcorollary}[Amortized reference dimension]
\label{cor:amortized-reference}
Let $\N,\M:A\to B$ be quantum channels with
$|A|=d_A$, $|B|=d_B\ge2$, and
$J_{\M}\ge\kappa I_{d_A d_B}/d_B$ for some $0<\kappa\le1$.
For $0<\varepsilon\le1$, let $m$ be the least
integer satisfying \eqref{eq:physical-explicit-length},
and set
\begin{align}\label{eq:amortized-reference-dimension}
r_m&:=\left\lceil d_A\sqrt{
\binom{m+(d_A d_B)^2-2}{(d_A d_B)^2-1}
\binom{m+d_A^4-2}{d_A^4-1}}
\right\rceil.
\end{align}
Then
\begin{align}\label{eq:amortized-reference-error}
0&\le D^{A}(\N\|\M)
-D^{A,(r_m)}(\N\|\M)<\varepsilon.
\end{align}
In particular, $r_m$ is polynomially bounded in
$1/\varepsilon$ for fixed $d_A,d_B,\kappa$.
\end{boxcorollary}

\begin{proof}
By concavity of the objective in
\eqref{eq:input-objective}, averaging an optimal
$m$-copy marginal over input permutations gives
a permutation-invariant optimizer $\tau$.
Its canonical purification $\psi$ is invariant
under simultaneous permutations of reference
and input pairs.

For integers $k\ge1$ and $n\ge1$, let $\mathfrak A_k(n)$
denote the algebra of operators on $(\mathbb C^k)^{\otimes n}$
that are invariant under conjugation by all unitaries that permute
the tensor factors, and set $\mathfrak A_k(0)=\mathbb C$.
Its dimension is
\begin{align}\label{eq:permutation-algebra-dimension}
\dim\mathfrak A_k(n)
&=\binom{n+k^2-1}{k^2-1},\qquad n\ge0.
\end{align}

For $0\le j\le m$, define
$\rho_j:=(\id_R\otimes\N^{\otimes j}
\otimes\id_{A^{\otimes(m-j)}})(\psi)$,
and $\sigma_j$ analogously with $\M$.
Since $J_{\M}>0$, there exists $c\ge1$ such that
$J_{\N}\le cJ_{\M}$. Complete positivity of $c\M-\N$
then gives $\rho_j\le c^j\sigma_j$ for every $0\le j\le m$.
In particular, all relative entropies in the telescoping sum below
are finite.

Before the $j$th use, the reference consists of
$R_j$, the processed pairs $R_iB_i$ ($i<j$),
and the unprocessed pairs $R_iA_i$ ($i>j$).
Its permutation-invariant algebra is
$\mathcal B(\mathbb C^{d_A})\otimes\mathfrak A_{d_A d_B}(j-1)
\otimes\mathfrak A_{d_A^2}(m-j)$.
By Schur--Weyl decomposition \cite[Section~III]{BP26},
each simple matrix block of $\mathfrak A_k(n)$ has size at most
$\sqrt{\dim\mathfrak A_k(n)}$, since the algebra dimension is
the sum of the squared block sizes.
The reference-algebra blocks therefore have size at most
$d_A\sqrt{\dim\mathfrak A_{d_A d_B}(j-1)
\dim\mathfrak A_{d_A^2}(m-j)}\le r_m$,
where the last inequality follows from
\eqref{eq:permutation-algebra-dimension} and $j-1,m-j\le m-1$.

For each $1\le j\le m$, a common unitary change of basis on the
reference gives the decompositions
\begin{align}\label{eq:amortized-block-decomposition}
\rho_{j-1}
&=\bigoplus_{\lambda}p_{j,\lambda}\,
\rho_{j,\lambda}\otimes\omega_{j,\lambda},\notag\\
\sigma_{j-1}
&=\bigoplus_{\lambda}q_{j,\lambda}\,
\sigma_{j,\lambda}\otimes\omega_{j,\lambda}.
\end{align}
Here $p_j$ and $q_j$ are the respective block-weight distributions,
and $\omega_{j,\lambda}$ is the normalized identity on the
multiplicity space. The normalized states
$\rho_{j,\lambda},\sigma_{j,\lambda}$ act on
$R_{j,\lambda}\otimes A_j$, where $R_{j,\lambda}$ is the
simple reference factor and $|R_{j,\lambda}|\le r_m$;
zero-weight terms are omitted.
The common multiplicity factor can be removed without changing
relative entropy, both before and after the channel use.
Moreover, $\rho_{j-1}\le c^{j-1}\sigma_{j-1}$ ensures that
$p_{j,\lambda}>0$ implies $q_{j,\lambda}>0$ and
$D(\rho_{j,\lambda}\|\sigma_{j,\lambda})<\infty$.
As $\rho_0=\sigma_0=\psi$,
we obtain
\begin{align}
D(\N^{\otimes m}\|\M^{\otimes m})
&=\sum_{j=1}^m
[D(\rho_j\|\sigma_j)-D(\rho_{j-1}\|\sigma_{j-1})]\notag\\
&=\sum_{j=1}^m\sum_{\lambda:p_{j,\lambda}>0}p_{j,\lambda}\Bigl[
D\bigl((\id\otimes\N)(\rho_{j,\lambda})
\big\|(\id\otimes\M)(\sigma_{j,\lambda})\bigr)
-D(\rho_{j,\lambda}\|\sigma_{j,\lambda})
\Bigr]\notag\\
&\le mD^{A,(r_m)}(\N\|\M)
\le mD^{\mathrm{reg}}(\N\|\M),\label{eq:amortized-telescope}
\end{align}
where the second equality follows from the direct-sum formula
for relative entropy: the channel acts only on $A_j$ and preserves
each state's block weights, so the classical term $D(p_j\|q_j)$
appears both before and after the channel use and cancels.
The first inequality holds because every block pair with
$p_{j,\lambda}>0$ is admissible in the definition of
$D^{A,(r_m)}(\N\|\M)$.
The last inequality follows from
$D^{A,(r_m)}(\N\|\M)\le D^A(\N\|\M)
=D^{\mathrm{reg}}(\N\|\M)$
\cite[Corollary~3]{FFRS}.
Dividing by $m$ and applying Theorem~\ref{thm:physical-block}
proves the claim.
\end{proof}

\subsection{Certified upper approximation}
\label{subsec:upper-approximation}

Theorem~\ref{thm:physical-block} bounds the blocklength
needed to approximate $D^{\mathrm{reg}}(\N\|\M)$ from below.
We now obtain an efficiently computable upper certificate using
the sharp R\'enyi hierarchy.
Combining~\cite{FST22} with Lemma~\ref{lem:renyi-continuity}
gives accuracy $\varepsilon$ at hierarchy level
$\widetilde O(\varepsilon^{-2})$.
This level bound concerns a different approximation sequence and
does not specify the blocklength required by
Theorem~\ref{thm:physical-block}.

For $\alpha>1$ and channel pairs with $J_{\M}>0$, the sharp R\'enyi channel divergence
\cite[Theorem~4.1]{FF21} is defined by
\begin{align}\label{eq:sharp-renyi-channel}
D^\#_\alpha(\N\|\M)
&:=\frac1{\alpha-1}\ln
\min_{\substack{W\ge0\\
J_{\N}\le J_{\M}\#_{1/\alpha}W}}
\|\Tr_BW\|_\infty,
\end{align}
where $X\#_tY=X^{1/2}(X^{-1/2}YX^{-1/2})^tX^{1/2}$
denotes the weighted matrix geometric mean for positive definite $X$,
positive semidefinite $Y$, and $0<t<1$.
Its finite-level error and efficient symmetry reduction
are established in \cite[Theorems~4.1, 4.3, 4.6]{FST22}.
Combining this error estimate with Lemma~\ref{lem:renyi-continuity}
gives a certified upper approximation to the Umegaki regularized
value.

\begin{boxtheorem}[Certified upper approximation]
\label{thm:upper-hierarchy}
Let $\N,\M:A\to B$ be quantum channels with
$|A|=d_A$, $|B|=d_B\ge2$, and
$J_{\M}\ge\kappa I_{d_A d_B}/d_B$ for some $0<\kappa\le1$.
Set $C':=2^{11}[(d_A d_B)^2+d_A d_B]$.
For every $0<\varepsilon\le1$, the following hold.
\begin{enumerate}[label=\textnormal{(\roman*)},ref=(\roman*)]
\item For every integer
\begin{align}\label{eq:upper-explicit-length}
m\ge\frac{C'}{\varepsilon^2}
\left[1+\ln\frac{d_B^2}{\kappa\varepsilon}\right]^2
\ln\left(
d_A d_B+\frac{C'}{\varepsilon^2}
\left[1+\ln\frac{d_B^2}{\kappa\varepsilon}\right]^2
\right),
\end{align}
there is an explicitly specified rational order
$\alpha=1/(1-2^{-\ell})$, with an integer $\ell\ge1$, such that
\begin{align}\label{eq:upper-hierarchy-error}
0&\le\frac1mD^\#_\alpha
(\N^{\otimes m}\|\M^{\otimes m})
-D^{\mathrm{reg}}(\N\|\M)
\le\frac{3\varepsilon}{4}.
\end{align}

\item If the Choi matrices are rational with total
encoding length $L$, an interval of width at most $\varepsilon$
containing $D^{\mathrm{reg}}(\N\|\M)$ can be
computed in $\operatorname{poly}(L,1/\varepsilon)$
time for fixed $d_A,d_B$. The algorithm requires
a hierarchy level $\widetilde O(\varepsilon^{-2})$
and an SDP lift of logarithmic depth for fixed
$d_A,d_B,\kappa$.
\end{enumerate}
\end{boxtheorem}

\begin{proof}
\emph{(i) Approximation error.}
Choose
\begin{align}
\ell=\left\lceil\log\left(
\frac{128}{\varepsilon}
[1+\ln(d_B^2/(\kappa\varepsilon))]^2
\right)\right\rceil,
\qquad s=2^{-\ell},
\qquad \alpha=\frac1{1-s}.
\end{align}
By \cite[Theorem~4.1]{FST22} and
\eqref{eq:upper-explicit-length},
\begin{align}\label{eq:sharp-hierarchy-bound}
0&\le\frac1mD^\#_\alpha
(\N^{\otimes m}\|\M^{\otimes m})
-\widetilde D_\alpha^{\mathrm{reg}}(\N\|\M)\notag\\
&\le\frac{[(d_A d_B)^2+d_A d_B]\ln(m+d_A d_B)}{ms}
\le\frac{\varepsilon}{4}.
\end{align}
Moreover, Lemma~\ref{lem:renyi-continuity}
gives $\widetilde D_\alpha^{\mathrm{reg}}(\N\|\M)
-D^{\mathrm{reg}}(\N\|\M)\le\varepsilon/2$.
Adding these bounds proves
\eqref{eq:upper-hierarchy-error}.

\emph{(ii) Certified computation.}
By \cite[Theorem~4.1]{FF21}, the sharp
R\'enyi channel divergence admits an SDP
representation for rational orders.
For $1/\alpha=1-2^{-\ell}$, the weighted
geometric mean has an SDP lift of depth $\ell$.
The symmetry reduction and efficient construction
of \cite[Theorems~4.3 and~4.6]{FST22}
therefore yield an SDP of size
$\operatorname{poly}(m,\ell)$ for fixed $d_A,d_B$.

For rational Choi matrices of total bit length
$L$, the symmetry-reduced SDP has coefficients
of bit length $\operatorname{poly}(m,L,\ell)$.
Since $J_{\M}>0$, an explicit strictly feasible
point and a bounded optimal region can be chosen
with polynomially controlled logarithmic radii.
Standard certified SDP optimization \cite{GLS93}
then computes $m^{-1}D^\#_\alpha(\N^{\otimes m}\|\M^{\otimes m})$ to accuracy
$\varepsilon/4$ in
$\operatorname{poly}(m,L,\ell,\log(1/\varepsilon))$
time.

Since $\log(1/\kappa)=\operatorname{poly}(L)$
for rational $J_{\M}>0$, choosing the smallest
admissible $m$ gives
$m=\operatorname{poly}(L,1/\varepsilon)$ and
$\ell=\operatorname{poly}(L,\log(1/\varepsilon))$.
Together with part~(i), the computed interval
yields an $\varepsilon$-accurate certificate for
$D^{\mathrm{reg}}(\N\|\M)$ in
$\operatorname{poly}(L,1/\varepsilon)$ time.
\end{proof}

The explicit convergence bound in Theorem~\ref{thm:physical-block}
reduces approximation of $D^{\mathrm{reg}}(\N\|\M)$ to evaluating
$D(\N^{\otimes m}\|\M^{\otimes m})$ at a sufficiently large blocklength $m$.
This reduction suggests two alternative approaches to efficient approximation.
The first would discretize Frenkel's integral representation of quantum
relative entropy to formulate a semidefinite approximation of the channel
optimization~\cite{Frenkel23,KW26}.
The second would combine numerical evaluation via matrix-logarithm
approximations~\cite{FSP19} with convex optimization, exploiting the
objective's concavity in the input marginal.

\subsection{Uniform continuity and computable-input extension}
\label{sec:uniform-stability}

In this subsection, we extend the certified approximation guarantees
from rational Choi matrices to computable Choi matrices, whose entries
may be irrational.
A Choi matrix $J$ is \emph{computable} if an algorithm, given any
integer $p\ge1$, returns a Hermitian matrix $H_p$ whose entries have
rational real and imaginary parts and which satisfies
$\|H_p-J\|_\infty\le2^{-p}$.
To control the resulting approximation error, we establish a
continuity bound for the finite-block divergence rates that is
uniform in blocklength and hence also applies to the regularized
divergence. The proof combines the successive channel replacement method
\cite[proof of Theorem~11]{LS09} with the
equal-marginal conditional-entropy bound
\cite{BLT25}, which improves upon
\cite[Lemma~2]{Winter16}.

\begin{boxproposition}[Uniform continuity]
\label{prop:uniform-stability}
Let $\N,\M,\N',\M':A\to B$ be quantum channels
with $|A|=d_A$, $|B|=d_B\ge2$, and
$J_{\M}\ge\kappa I_{d_A d_B}/d_B$ for some $0<\kappa\le1$.
Fix $0\le\delta\le\kappa/(2d_B)$.
Define $\omega_\kappa(0):=0$ and, for $\delta>0$,
\begin{align}\label{eq:stability-modulus}
\omega_\kappa(\delta):=\frac{d_B\delta}{2}
\left(2+\ln\frac{2d_B^3}{\kappa\delta}\right)
+\frac{2d_B\delta}{\kappa}.
\end{align}
If $\|J_{\N}-J_{\N'}\|_\infty\le\delta$ and
$\|J_{\M}-J_{\M'}\|_\infty\le\delta$,
then, for every $m\ge1$,
\begin{align}\label{eq:uniform-stability}
\left|
\frac{D(\N^{\otimes m}\|\M^{\otimes m})}{m}
-\frac{D(\N'^{\otimes m}\|\M'^{\otimes m})}{m}
\right|&\le\omega_\kappa(\delta),
\end{align}
and
\begin{align}\label{eq:regularized-stability}
\left|D^{\mathrm{reg}}(\N\|\M)
-D^{\mathrm{reg}}(\N'\|\M')\right|
&\le\omega_\kappa(\delta).
\end{align}
\end{boxproposition}

\begin{proof}
The case $\delta=0$ is immediate.
Set $t:=d_B\delta/2\le\kappa/4$.
Fix an $m$-copy input and successively replace
$\N$ by $\N'$ in each channel use, keeping
$\M^{\otimes m}$ fixed.
Let $B_i$ denote the output of the $i$-th use
and $Z$ all remaining systems, including the reference.
Since $-\delta I_{d_A d_B}\le J_{\N}-J_{\N'}\le\delta I_{d_A d_B}$,
we have
\begin{align}
-d_B\delta\mathcal R
\le_{\mathrm{CP}}\N-\N'
\le_{\mathrm{CP}}d_B\delta\mathcal R,
\end{align}
where $\mathcal R(X)=\Tr(X)I_{d_B}/d_B$.
Each replacement changes the output by trace
distance at most $t$, while preserving its marginal on $Z$.

Let $\sigma$ be the fixed output of $\M^{\otimes m}$.
CP domination and pinching give
\begin{align}\label{eq:stability-spectral-bound}
\frac{\kappa}{d_B}I_{B_i}\otimes\sigma_Z
\le\sigma\le d_B I_{B_i}\otimes\sigma_Z.
\end{align}
Thus $\ln\sigma-\ln(I_{B_i}\otimes\sigma_Z)$
has spectral width at most $\ln(d_B^2/\kappa)$
on the common support of $\sigma$ and $I_{B_i}\otimes\sigma_Z$.

For consecutive outputs $\rho,\rho'$ with
$\rho_Z=\rho'_Z$, we have
\begin{align}
D(\rho\|\sigma)-D(\rho'\|\sigma)
&=H(B_i|Z)_{\rho'}-H(B_i|Z)_\rho
-\Tr(\rho-\rho')
\bigl[\ln\sigma-\ln(I_{B_i}\otimes\sigma_Z)\bigr],\notag\\
|D(\rho\|\sigma)-D(\rho'\|\sigma)|
&\le t\ln\frac{d_B^2(d_B^2-1)}{\kappa}+h_2(t)
\le t\left(2+\ln\frac{d_B^4}{\kappa t}\right).
\end{align}
Here $H(B_i|Z)_\rho:=
-\Tr\rho\ln\rho+\Tr\rho_Z\ln\rho_Z$
is the conditional von Neumann entropy, and
$h_2(t):=-t\ln t-(1-t)\ln(1-t)$ is the binary entropy in nats.
The inequality follows from
the equal-marginal continuity bound
\cite[Theorem~5]{BLT25}
and the spectral-width estimate.

For the change from $\M^{\otimes m}$ to
$(\M')^{\otimes m}$, the Choi bound gives
\begin{align}
(1-d_B\delta/\kappa)\M
\le_{\mathrm{CP}}\M'
\le_{\mathrm{CP}}(1+d_B\delta/\kappa)\M.
\end{align}
The corresponding $m$-copy outputs satisfy
$(1-d_B\delta/\kappa)^m\sigma
\le\sigma'\le(1+d_B\delta/\kappa)^m\sigma$.
Since $d_B\delta/\kappa\le1/2$, the states $\sigma$ and $\sigma'$
have the same support. Operator monotonicity of the logarithm on
this common support gives
\begin{align}
|D(\rho\|\sigma)-D(\rho\|\sigma')|
&=|\Tr\rho(\ln\sigma'-\ln\sigma)|
\le-m\ln(1-d_B\delta/\kappa)
\le\frac{2m d_B\delta}{\kappa}.
\end{align}

Summing the $m$ errors from replacing $\N$
and adding the bound for replacing $\M$,
then maximizing over inputs, gives
\eqref{eq:uniform-stability}.
Taking suprema over $m$ proves
\eqref{eq:regularized-stability}.
\end{proof}

A rational approximation of a Choi matrix need not
define a quantum channel. For computable Choi matrices,
we correct the partial trace and restore positive semidefiniteness,
then apply Proposition~\ref{prop:uniform-stability}.

\begin{boxcorollary}[Computable-input extension]
\label{cor:real-input}
Let $\N,\M:A\to B$ be quantum channels with
$|A|=d_A$, $|B|=d_B\ge2$, and
$J_{\M}\ge\kappa I_{d_A d_B}/d_B$ for a known rational
$0<\kappa\le1$.
Suppose $J_{\N}$ and $J_{\M}$ are computable,
in the sense that certified rational approximations
can be obtained to arbitrary operator-norm precision.
Then the certified approximation guarantees of
Theorem~\ref{thm:upper-hierarchy}
extend to these channels.
For fixed $d_A,d_B$, accuracy $0<\varepsilon\le1$
requires only
$O(\log(1/\kappa)+\log(1/\varepsilon))$
bits of input precision.
\end{boxcorollary}

\begin{proof}
Choose a positive dyadic rational $\varepsilon_1$
such that
\begin{align}\label{eq:real-input-precision}
\frac{\kappa\varepsilon^2}{256d_B^5}
<\varepsilon_1
\le\frac{\kappa\varepsilon^2}{128d_B^5}.
\end{align}
For each Choi matrix $J$, obtain a rational Hermitian
approximation $H$ with
$\|H-J\|_\infty\le\varepsilon_1$, and define
\begin{align}\label{eq:choi-rational-repair}
T&:=H+(I_{d_A}-\Tr_BH)\otimes I_{d_B}/d_B,\qquad
\widehat J:=\frac{T+2\varepsilon_1I_{d_A d_B}}
{1+2d_B\varepsilon_1}.
\end{align}
Since $\Tr_BT=I_{d_A}$ and
$\|T-J\|_\infty\le2\varepsilon_1$,
we have $\widehat J\ge0$ and
$\Tr_B\widehat J=I_{d_A}$.
Thus $\widehat J$ is a rational channel Choi matrix
\cite[Corollary~2.27]{Watrous18}.

Moreover,
$0\le T+2\varepsilon_1I_{d_A d_B}-J
\le4\varepsilon_1I_{d_A d_B}$.
Since $0\le J\le d_B I_{d_A d_B}$, it follows that
\begin{align}\label{eq:choi-repair-error}
\|\widehat J-J\|_\infty
&\le2d_B^2\varepsilon_1,\qquad
\widehat J\ge\frac{J}{1+2d_B\varepsilon_1}.
\end{align}

Apply the construction to $J_{\N}$ and $J_{\M}$.
Set $\delta:=\kappa\varepsilon^2/(64d_B^3)$.
By \eqref{eq:choi-repair-error}, both repaired
matrices are within $\delta$ of the originals.
In particular,
$\widehat J_{\M}\ge J_{\M}/(1+2d_B\varepsilon_1)
\ge\kappa I_{d_A d_B}/(2d_B)$.
Since $\delta\le\kappa/(2d_B)$,
Proposition~\ref{prop:uniform-stability} gives
\begin{align}
\omega_\kappa(\delta)
&\le\frac{\varepsilon}{128d_B^2}
\left(2+\ln128+6\ln d_B+\frac2e\right)
+\frac{\varepsilon}{32d_B^2}
<\frac{\varepsilon}{8},
\end{align}
where we used $x\ln(1/x)\le1/e$
for $0<x\le1$.

Apply the rational-input algorithms of
Theorem~\ref{thm:upper-hierarchy}
with certified interval width at most $\varepsilon/2$.
Enlarging each interval by $\omega_\kappa(\delta)$
at both endpoints gives an interval for the original
channels of width less than
$\varepsilon/2+2(\varepsilon/8)<\varepsilon$.

Finally, since
$\varepsilon_1=\Theta(\kappa\varepsilon^2/d_B^5)$
by \eqref{eq:real-input-precision},
the rational approximations require only
$O(\log(1/\kappa)+\log(1/\varepsilon))$
bits of precision for fixed $d_A,d_B$.
\end{proof}

\section{Conclusion}\label{sec:conclusion}

We established a sharp finite-attainment dichotomy for regularized
quantum channel relative entropy: the regularized value is attained
at one copy or remains unattained at every finite blocklength.
We developed two-copy criteria for distinguishing these cases and
showed that interiority, strict superadditivity, and finite nonattainment
are generic.
On the computational side, we established a finite-block approximation bound
and developed a semidefinite hierarchy that provides efficient,
certified approximation of the regularized value.
Together, these results clarify the structure, generic behavior, and
computational accessibility of this fundamental quantity.

Several questions remain open. One problem is to
identify additional conditions under which two-copy additivity
characterizes finite attainment, particularly for higher-dimensional
channels with rank-deficient optimizers.
Extending these criteria to other channel divergences is another
natural direction.
Another problem is to determine the optimal finite-block
convergence rate by closing the gap between the necessary
$\Omega(\varepsilon^{-1})$ and sufficient
$\widetilde O(\varepsilon^{-3/2})$ blocklength bounds mentioned in
Remark~\ref{rem:block-lower-bound}.

\section*{Acknowledgment}

K.F. and Y.X. are supported in part by the National Natural Science Foundation of China (Grant Nos. 92470113 and 12404569), the Shenzhen Science and Technology Program (Grant Nos. QNXMB20250701091826036 and JCYJ20240813113519025), the Shenzhen Fundamental Research Program (Grant No. JCYJ20241202124023031),  the Guangdong Provincial Quantum Science Strategic Initiative (Grant Nos. GDZX2603008 and GDZX2503001) and the General R\&D Projects of 1+1+1 CUHK-CUHK(SZ)-GDST Joint Collaboration Fund (Grant No. GRDP2025-022). 
We acknowledge that OpenAI GPT-6 Astra was used to explore proof strategies and support the writing of the manuscript. We have verified AI-assisted material and take full responsibility for the content of this work.

\bibliographystyle{alphaurl}
\bibliography{ref}

\newcommand{\etalchar}[1]{$^{#1}$}
\begin{thebibliography}{SPGWC10}

\bibitem[BLT25]{BLT25}
Mario Berta, Ludovico Lami, and Marco Tomamichel.
\newblock Continuity of entropies via integral representations.
\newblock {\em IEEE Transactions on Information Theory}, 71(3):1896--1908, 2025.
\newblock \href {https://arxiv.org/abs/2408.15226} {\path{arXiv:2408.15226}}, \href {https://doi.org/10.1109/TIT.2025.3527858} {\path{doi:10.1109/TIT.2025.3527858}}.

\bibitem[BP26]{BP26}
Bjarne Bergh and Marco Parentin.
\newblock Permutation invariant optimization problems in quantum information theory: A framework for channel fidelity and beyond, 2026.
\newblock \href {https://arxiv.org/abs/2604.27040} {\path{arXiv:2604.27040}}.

\bibitem[CG19]{CG19}
Eric Chitambar and Gilad Gour.
\newblock Quantum resource theories.
\newblock {\em Reviews of Modern Physics}, 91(2):025001, 2019.
\newblock \href {https://arxiv.org/abs/1806.06107} {\path{arXiv:1806.06107}}, \href {https://doi.org/10.1103/RevModPhys.91.025001} {\path{doi:10.1103/RevModPhys.91.025001}}.

\bibitem[CML16]{CML16}
Patrick~J. Coles, Eric~M. Metodiev, and Norbert L{\"u}tkenhaus.
\newblock Numerical approach for unstructured quantum key distribution.
\newblock {\em Nature Communications}, 7:11712, 2016.
\newblock \href {https://doi.org/10.1038/ncomms11712} {\path{doi:10.1038/ncomms11712}}.

\bibitem[CMW16]{CMW16}
T.~Cooney, M.~Mosonyi, and M.~M. Wilde.
\newblock Strong converse exponents for a quantum channel discrimination problem and quantum-feedback-assisted communication.
\newblock {\em Communications in Mathematical Physics}, 344:797--829, 2016.
\newblock \href {https://arxiv.org/abs/1408.3373} {\path{arXiv:1408.3373}}.

\bibitem[FF21a]{FF21GRD}
Kun Fang and Hamza Fawzi.
\newblock Geometric {R\'enyi} divergence and its applications in quantum channel capacities.
\newblock {\em Communications in Mathematical Physics}, 384:1615--1677, 2021.
\newblock \href {https://arxiv.org/abs/1909.05758} {\path{arXiv:1909.05758}}, \href {https://doi.org/10.1007/s00220-021-04064-4} {\path{doi:10.1007/s00220-021-04064-4}}.

\bibitem[FF21b]{FF21}
Hamza Fawzi and Omar Fawzi.
\newblock Defining quantum divergences via convex optimization.
\newblock {\em Quantum}, 5:387, 2021.
\newblock \href {https://arxiv.org/abs/2007.12576} {\path{arXiv:2007.12576}}, \href {https://doi.org/10.22331/q-2021-01-26-387} {\path{doi:10.22331/q-2021-01-26-387}}.

\bibitem[FFRS20]{FFRS}
Kun Fang, Omar Fawzi, Renato Renner, and David Sutter.
\newblock A chain rule for the quantum relative entropy.
\newblock {\em Physical Review Letters}, 124:100501, 2020.
\newblock \href {https://arxiv.org/abs/1909.05826} {\path{arXiv:1909.05826}}, \href {https://doi.org/10.1103/PhysRevLett.124.100501} {\path{doi:10.1103/PhysRevLett.124.100501}}.

\bibitem[Fre23]{Frenkel23}
P{\'e}ter~E. Frenkel.
\newblock Integral formula for quantum relative entropy implies data processing inequality.
\newblock {\em Quantum}, 7:1102, 2023.
\newblock \href {https://doi.org/10.22331/q-2023-09-07-1102} {\path{doi:10.22331/q-2023-09-07-1102}}.

\bibitem[FSP19]{FSP19}
Hamza Fawzi, James Saunderson, and Pablo~A. Parrilo.
\newblock Semidefinite approximations of the matrix logarithm.
\newblock {\em Foundations of Computational Mathematics}, 19:259--296, 2019.
\newblock \href {https://arxiv.org/abs/1705.00812} {\path{arXiv:1705.00812}}, \href {https://doi.org/10.1007/s10208-018-9385-0} {\path{doi:10.1007/s10208-018-9385-0}}.

\bibitem[FST22]{FST22}
Omar Fawzi, Ala Shayeghi, and Hoang Ta.
\newblock A hierarchy of efficient bounds on quantum capacities exploiting symmetry.
\newblock {\em IEEE Transactions on Information Theory}, 68(11):7346--7360, 2022.
\newblock \href {https://arxiv.org/abs/2203.02127} {\path{arXiv:2203.02127}}, \href {https://doi.org/10.1109/TIT.2022.3182101} {\path{doi:10.1109/TIT.2022.3182101}}.

\bibitem[GJL26]{GJL26}
Minbo Gao, Zhengfeng Ji, and Chenghua Liu.
\newblock Quantum channel {Stein's} lemma with an exponential strong converse, 2026.
\newblock \href {https://arxiv.org/abs/2609.27196} {\path{arXiv:2609.27196}}.

\bibitem[GLS93]{GLS93}
Martin Gr{\"o}tschel, L{\'a}szl{\'o} Lov{\'a}sz, and Alexander Schrijver.
\newblock {\em Geometric Algorithms and Combinatorial Optimization}.
\newblock Springer, 2 edition, 1993.
\newblock \href {https://doi.org/10.1007/978-3-642-78240-4} {\path{doi:10.1007/978-3-642-78240-4}}.

\bibitem[Hay17]{Hayashi17}
Masahito Hayashi.
\newblock {\em Quantum Information Theory: Mathematical Foundation}.
\newblock Graduate Texts in Physics. Springer, Berlin, Heidelberg, 2 edition, 2017.
\newblock \href {https://doi.org/10.1007/978-3-662-49725-8} {\path{doi:10.1007/978-3-662-49725-8}}.

\bibitem[HP91]{HP91}
F.~Hiai and D.~Petz.
\newblock The proper formula for relative entropy and its asymptotics in quantum probability.
\newblock {\em Communications in Mathematical Physics}, 143:99--114, 1991.
\newblock \href {https://doi.org/10.1007/BF02100287} {\path{doi:10.1007/BF02100287}}.

\bibitem[HP14]{HP14}
Fumio Hiai and D{\'e}nes Petz.
\newblock {\em Introduction to Matrix Analysis and Applications}.
\newblock Universitext. Springer, Cham, 2014.
\newblock \href {https://doi.org/10.1007/978-3-319-04150-6} {\path{doi:10.1007/978-3-319-04150-6}}.

\bibitem[JR18]{JaquesRahaman2018}
Samuel Jaques and Mizanur Rahaman.
\newblock Spectral properties of tensor products of channels.
\newblock {\em Journal of Mathematical Analysis and Applications}, 465(2):1134--1158, 2018.
\newblock \href {https://doi.org/10.1016/j.jmaa.2018.05.052} {\path{doi:10.1016/j.jmaa.2018.05.052}}.

\bibitem[KW26]{KW26}
Gereon Ko{\ss}mann and Mark~M. Wilde.
\newblock Semidefinite optimization of the quantum relative entropy of channels.
\newblock {\em IEEE Transactions on Information Theory}, 72(4):2378--2390, 2026.
\newblock \href {https://arxiv.org/abs/2410.16362} {\path{arXiv:2410.16362}}, \href {https://doi.org/10.1109/TIT.2025.3642733} {\path{doi:10.1109/TIT.2025.3642733}}.

\bibitem[LS09]{LS09}
Debbie Leung and Graeme Smith.
\newblock Continuity of quantum channel capacities.
\newblock {\em Communications in Mathematical Physics}, 292:201--215, 2009.
\newblock \href {https://arxiv.org/abs/0810.4931} {\path{arXiv:0810.4931}}, \href {https://doi.org/10.1007/s00220-009-0833-1} {\path{doi:10.1007/s00220-009-0833-1}}.

\bibitem[Mit20]{Mityagin20}
Boris~S. Mityagin.
\newblock The zero set of a real analytic function.
\newblock {\em Mathematical Notes}, 107:529--530, 2020.
\newblock \href {https://arxiv.org/abs/1512.07276} {\path{arXiv:1512.07276}}, \href {https://doi.org/10.1134/S0001434620030189} {\path{doi:10.1134/S0001434620030189}}.

\bibitem[MLDS{\etalchar{+}}13]{MLDS13}
Martin M{\"u}ller-Lennert, Fr{\'e}d{\'e}ric Dupuis, Oleg Szehr, Serge Fehr, and Marco Tomamichel.
\newblock On quantum {R}{\'e}nyi entropies: A new generalization and some properties.
\newblock {\em Journal of Mathematical Physics}, 54:122203, 2013.
\newblock \href {https://arxiv.org/abs/1306.3142} {\path{arXiv:1306.3142}}, \href {https://doi.org/10.1063/1.4838856} {\path{doi:10.1063/1.4838856}}.

\bibitem[NC10]{NC10}
Michael~A. Nielsen and Isaac~L. Chuang.
\newblock {\em Quantum Computation and Quantum Information}.
\newblock Cambridge University Press, 10th anniversary edition, 2010.
\newblock \href {https://doi.org/10.1017/CBO9780511976667} {\path{doi:10.1017/CBO9780511976667}}.

\bibitem[ON00]{ON00}
T.~Ogawa and H.~Nagaoka.
\newblock Strong converse and {Stein's} lemma in quantum hypothesis testing.
\newblock {\em IEEE Transactions on Information Theory}, 46:2428--2433, 2000.
\newblock \href {https://arxiv.org/abs/quant-ph/9906090} {\path{arXiv:quant-ph/9906090}}, \href {https://doi.org/10.1109/18.887855} {\path{doi:10.1109/18.887855}}.

\bibitem[Pet07]{Petz07}
D{\'e}nes Petz.
\newblock Bregman divergence as relative operator entropy.
\newblock {\em Acta Mathematica Hungarica}, 116:127--131, 2007.
\newblock \href {https://doi.org/10.1007/s10474-007-6014-9} {\path{doi:10.1007/s10474-007-6014-9}}.

\bibitem[SPGWC10]{Sanz2010Wielandt}
Mikel Sanz, David P{\'e}rez-Garc{\'i}a, Michael~M. Wolf, and Juan~I. Cirac.
\newblock A quantum version of {Wielandt}'s inequality.
\newblock {\em IEEE Transactions on Information Theory}, 56(9):4668--4673, 2010.
\newblock \href {https://doi.org/10.1109/TIT.2010.2054552} {\path{doi:10.1109/TIT.2010.2054552}}.

\bibitem[SS26]{SS26}
Lukas Schmitt and David Sutter.
\newblock The regularized channel {R\'enyi} divergence is continuous, 2026.
\newblock \href {https://arxiv.org/abs/2609.29885} {\path{arXiv:2609.29885}}.

\bibitem[VA16]{VA16}
Sai Vinjanampathy and Janet Anders.
\newblock Quantum thermodynamics.
\newblock {\em Contemporary Physics}, 57(4):545--579, 2016.
\newblock \href {https://arxiv.org/abs/1508.06099} {\path{arXiv:1508.06099}}, \href {https://doi.org/10.1080/00107514.2016.1201896} {\path{doi:10.1080/00107514.2016.1201896}}.

\bibitem[Ved02]{Vedral02}
Vlatko Vedral.
\newblock The role of relative entropy in quantum information theory.
\newblock {\em Reviews of Modern Physics}, 74(1):197--234, 2002.
\newblock \href {https://arxiv.org/abs/quant-ph/0102094} {\path{arXiv:quant-ph/0102094}}, \href {https://doi.org/10.1103/RevModPhys.74.197} {\path{doi:10.1103/RevModPhys.74.197}}.

\bibitem[Wat18]{Watrous18}
John Watrous.
\newblock {\em The Theory of Quantum Information}.
\newblock Cambridge University Press, 2018.
\newblock \href {https://doi.org/10.1017/9781316848142} {\path{doi:10.1017/9781316848142}}.

\bibitem[WBHK20]{WBHK20}
M.~M. Wilde, M.~Berta, C.~Hirche, and E.~Kaur.
\newblock Amortized channel divergence for asymptotic quantum channel discrimination.
\newblock {\em Letters in Mathematical Physics}, 110:2277--2336, 2020.
\newblock \href {https://arxiv.org/abs/1808.01498} {\path{arXiv:1808.01498}}.

\bibitem[WFD19]{WFD19}
Xin Wang, Kun Fang, and Runyao Duan.
\newblock Semidefinite programming converse bounds for quantum communication.
\newblock {\em IEEE Transactions on Information Theory}, 65(4):2583--2592, 2019.
\newblock \href {https://arxiv.org/abs/1709.00200} {\path{arXiv:1709.00200}}, \href {https://doi.org/10.1109/TIT.2018.2874031} {\path{doi:10.1109/TIT.2018.2874031}}.

\bibitem[WFT19]{WFT19}
Xin Wang, Kun Fang, and Marco Tomamichel.
\newblock On converse bounds for classical communication over quantum channels.
\newblock {\em IEEE Transactions on Information Theory}, 65(7):4609--4619, 2019.
\newblock \href {https://arxiv.org/abs/1709.05258} {\path{arXiv:1709.05258}}, \href {https://doi.org/10.1109/TIT.2019.2898656} {\path{doi:10.1109/TIT.2019.2898656}}.

\bibitem[Wil17]{Wilde17}
Mark~M. Wilde.
\newblock {\em Quantum Information Theory}.
\newblock Cambridge University Press, 2 edition, 2017.
\newblock \href {https://doi.org/10.1017/9781316809976} {\path{doi:10.1017/9781316809976}}.

\bibitem[Win16]{Winter16}
Andreas Winter.
\newblock Tight uniform continuity bounds for quantum entropies: conditional entropy, relative entropy distance and energy constraints.
\newblock {\em Communications in Mathematical Physics}, 347:291--313, 2016.
\newblock \href {https://arxiv.org/abs/1507.07775} {\path{arXiv:1507.07775}}, \href {https://doi.org/10.1007/s00220-016-2609-8} {\path{doi:10.1007/s00220-016-2609-8}}.

\bibitem[WLC18]{WLC18}
Adam Winick, Norbert L{\"u}tkenhaus, and Patrick~J. Coles.
\newblock Reliable numerical key rates for quantum key distribution.
\newblock {\em Quantum}, 2:77, 2018.
\newblock \href {https://doi.org/10.22331/q-2018-07-26-77} {\path{doi:10.22331/q-2018-07-26-77}}.

\bibitem[WW19]{WW19Distinguishability}
Xin Wang and Mark~M. Wilde.
\newblock Resource theory of asymmetric distinguishability for quantum channels.
\newblock {\em Physical Review Research}, 1:033169, 2019.
\newblock \href {https://arxiv.org/abs/1907.06306} {\path{arXiv:1907.06306}}, \href {https://doi.org/10.1103/PhysRevResearch.1.033169} {\path{doi:10.1103/PhysRevResearch.1.033169}}.

\bibitem[WWS19]{WWS}
Xin Wang, Mark~M. Wilde, and Yuan Su.
\newblock Quantifying the magic of quantum channels.
\newblock {\em New Journal of Physics}, 21:103002, 2019.
\newblock \href {https://arxiv.org/abs/1903.04483} {\path{arXiv:1903.04483}}.

\bibitem[WWY14]{WWY14}
Mark~M. Wilde, Andreas Winter, and Dong Yang.
\newblock Strong converse for the classical capacity of entanglement-breaking and hadamard channels via a sandwiched {R}{\'e}nyi relative entropy.
\newblock {\em Communications in Mathematical Physics}, 331:593--622, 2014.
\newblock \href {https://arxiv.org/abs/1306.1586} {\path{arXiv:1306.1586}}, \href {https://doi.org/10.1007/s00220-014-2122-x} {\path{doi:10.1007/s00220-014-2122-x}}.

\bibitem[WY26]{WY26}
Jinzhao Wang and Yuxiang Yang.
\newblock Continuity of regularized channel {R\'enyi} divergences, 2026.
\newblock \href {https://arxiv.org/abs/2609.28635} {\path{arXiv:2609.28635}}.

\bibitem[ZW26]{ZW26}
Chengkai Zhu and Xin Wang.
\newblock Quantum channel {Stein} theorem beyond definite causal order, 2026.
\newblock \href {https://arxiv.org/abs/2609.30268} {\path{arXiv:2609.30268}}.

\end{thebibliography}

\appendix
\section{Technical proofs}\label{app:boundary}

We derive the boundary perturbation estimates used in the proof of
Lemma~\ref{lem:qubit-boundary} and the chain-rule argument for
Theorem~\ref{thm:qubit}, followed by the general-dimensional alignment
Lemma~\ref{lem:general-block-alignment}.
In Sections~\ref{app:alignment} and~\ref{app:inward}, we assume
$0<D(\N\|\M)<\infty$ and that $\tau_0=|0\rangle\langle0|$
is a one-copy optimizer for the qubit channel pair.
We use the Choi block notation from the proof of Lemma~\ref{lem:qubit-boundary},
with $Z=M_0^{-1}Y$, $F=N_1-Z^\dagger N_0Z$, and
$G=M_1-Z^\dagger M_0Z$.

\subsection{Proof of block alignment}\label{app:alignment}

\begin{proof}[Proof of the alignment facts in \eqref{eq:qubit-block-alignment}]
Set $S=N_1-X^\dagger N_0^+X$.
By positive semidefiniteness of the Choi matrices and the Schur
complement criterion,
\begin{align}
\ran X\subseteq\ran N_0,\qquad
S\ge0,\qquad
G=M_1-Y^\dagger M_0^{-1}Y\ge0.
\end{align}
For $\tau_p=\operatorname{diag}(1-p,p)$, write
\begin{align}
\rho_p=\begin{pmatrix}
(1-p)N_0&\sqrt{p(1-p)}X\\
\sqrt{p(1-p)}X^\dagger&pN_1
\end{pmatrix},
\quad
\sigma_p=\begin{pmatrix}
(1-p)M_0&\sqrt{p(1-p)}Y\\
\sqrt{p(1-p)}Y^\dagger&pM_1
\end{pmatrix}.
\end{align}

Define
\begin{align}
R_p=\begin{pmatrix}\sqrt{1-p}\,I\\
\sqrt p\,X^\dagger N_0^+\end{pmatrix},
\qquad
L_p=\begin{pmatrix}\sqrt{1-p}\,I\\
\sqrt p\,Z^\dagger\end{pmatrix}.
\end{align}
Using $N_0N_0^+X=X$ and $M_0Z=Y$, direct multiplication gives
\begin{align}
\rho_p=R_pN_0R_p^\dagger+p(0\oplus S),
\qquad
\sigma_p=L_pM_0L_p^\dagger+p(0\oplus G).
\end{align}

As $p\to0$, we have
\begin{align}
R_p^\dagger R_p=(1-p)I+pN_0^+XX^\dagger N_0^+=I+O(p)>0.
\end{align}
The nonzero eigenvalues of $R_pN_0R_p^\dagger$ are
those of $N_0^{1/2}R_p^\dagger R_pN_0^{1/2}=N_0+O(p)$.
Since $R_p^\dagger R_p>0$, this matrix has the same
kernel as $N_0$, so zero eigenvalues remain zero and
\begin{align}
\Tr R_pN_0R_p^\dagger\ln(R_pN_0R_p^\dagger)
=\Tr N_0\ln N_0+O(p).
\end{align}
Also, $p\Tr S\ln(pS)=p\ln p\,\Tr S+O(p)$.
By operator monotonicity of the logarithm,
\begin{align}
\Tr\rho_p\ln\rho_p
\ge\Tr N_0\ln N_0+p\ln p\,\Tr S-O(p).
\end{align}

Choose $c>\|G\|$ and put
\begin{align}
M_p=L_pM_0L_p^\dagger+cpI,
\qquad
P_p=I-L_p(L_p^\dagger L_p)^{-1}L_p^\dagger.
\end{align}
Since $\sigma_p\le M_p$, operator monotonicity of logarithm gives $\Tr\rho_p\ln\sigma_p
\le\Tr\rho_p\ln M_p$.

Since $P_pL_p=0$ and
\begin{align}
(P_p)_{22}
=I-pZ^\dagger(L_p^\dagger L_p)^{-1}Z=I+O(p),
\qquad
R_p-L_p=\sqrt p
\begin{pmatrix}0\\(N_0^+X-Z)^\dagger\end{pmatrix},
\end{align}
we obtain
\begin{align}
\Tr\rho_pP_p
&=\Tr N_0(R_p-L_p)^\dagger P_p(R_p-L_p)
+p\Tr S(P_p)_{22}\notag\\
&=p\Tr\bigl[S+(N_0^+X-Z)^\dagger
N_0(N_0^+X-Z)\bigr]+O(p^2).
\end{align}

Since $P_pL_p=0$, we have
$\ran P_p=(\ran L_p)^\perp$. Moreover, direct multiplication gives
\begin{align}
L_p^\dagger L_p=I+O(p),\qquad
L_p^\dagger\rho_pL_p=N_0+O(p),\qquad
L_p^\dagger M_pL_p=M_0+O(p).
\end{align}
Thus the orthogonal compressions of $\rho_p$ and
$M_p$ to $\ran L_p$ are $N_0+O(p)$ and $M_0+O(p)$,
respectively, while $M_p=cpI$ on $\ran P_p$.
Since $M_0>0$, functional calculus gives
\begin{align}
\Tr\rho_p\ln\sigma_p
&\le\Tr\rho_p\ln M_p
=\Tr N_0\ln M_0+\ln(cp)\Tr\rho_pP_p+O(p)\notag\\
&=\Tr N_0\ln M_0+p\ln p\,
\Tr\bigl[S+(N_0^+X-Z)^\dagger N_0(N_0^+X-Z)\bigr]+O(p).
\end{align}

Consequently,
\begin{align}
D(\N,\M;\tau_p)
&=\Tr\rho_p\ln\rho_p-\Tr\rho_p\ln\sigma_p\notag\\
&\ge D(\N\|\M)+p\ln\frac1p\,
\Tr(N_0^+X-Z)^\dagger N_0(N_0^+X-Z)-O(p).\label{eq:log-boundary}
\end{align}
Since $D(\N,\M;\tau_p)\le D(\N\|\M)$ and
$p\ln(1/p)$ dominates $O(p)$ as $p\to0$,
we obtain
\begin{align}
\Tr(N_0^+X-Z)^\dagger N_0(N_0^+X-Z)=0.
\end{align}
As $N_0\ge0$, this implies $N_0(N_0^+X-Z)=0$.
Using $N_0N_0^+X=X$, we conclude $X=N_0Z$.

Finally, $F=N_1-Z^\dagger N_0Z=S\ge0$.
Compressing $J_{\N}\le\lambda J_{\M}$
by $\binom{-Z}{I}$ gives $F\le\lambda G$.
\end{proof}

\subsection{Coherent and inward perturbations}\label{app:inward}

\begin{proof}[Proof of the coherent trace condition \eqref{eq:qubit-coherent-trace}]
For $t\in\mathbb C$, let
\begin{align}
\tau_t=\frac1{1+|t|^2}
\begin{pmatrix}1&\bar t\\t&|t|^2\end{pmatrix},
\end{align}
which is a rank-one projection. Direct block
multiplication gives
\begin{align}
(\tau_t\otimes I)J_{\N}(\tau_t\otimes I)
=\tau_t\otimes
\frac{N_0+tX+\bar tX^\dagger+|t|^2N_1}{1+|t|^2},
\end{align}
and similarly for $\M$. By homogeneity
and the alignment identities in \eqref{eq:qubit-block-alignment},
\begin{align}
D(\N,\M;\tau_t)
&=\frac1{1+|t|^2}
D\Bigl((I+\bar tZ^\dagger)N_0(I+tZ)+|t|^2F
\Big\|(I+\bar tZ^\dagger)M_0(I+tZ)+|t|^2G\Bigr).
\end{align}

Let $L_t=I+tZ$ and $H_t=L_tL_t^\dagger=I+tZ+\bar tZ^\dagger+|t|^2ZZ^\dagger$,
and
\begin{align}
D_{N_0,M_0}(H_t)=D(\sqrt{H_t} N_0\sqrt{H_t}\|\sqrt{H_t} M_0\sqrt{H_t}).
\end{align}
By polar decomposition,
$D_{N_0,M_0}(H_t)=D(L_t^\dagger N_0L_t\|L_t^\dagger M_0L_t)$.
For $t=re^{i\theta}$, the derivative in
Lemma~\ref{lem:supporting-bound} gives
\begin{align}
\left.\frac{d}{dr}D_{N_0,M_0}(H_{re^{i\theta}})\right|_{r=0}
=\Tr(e^{i\theta}Z+e^{-i\theta}Z^\dagger)
\mathfrak B(N_0,M_0)
=2\Re\bigl(e^{i\theta}\Tr\mathfrak B(N_0,M_0)Z\bigr).
\end{align}
Here the quadratic term has zero first
derivative at $r=0$, and $\mathfrak B(N_0,M_0)$
is Hermitian. By smoothness of $D_{N_0,M_0}$ near $I$,
\begin{align}
D_{N_0,M_0}(H_t)&=D(\N\|\M)+
2\Re\bigl(t\Tr\mathfrak B(N_0,M_0)Z\bigr)
+O(|t|^2).
\end{align}

Since $M_0>0$, $L_t^\dagger M_0L_t\ge cI$
for some $c>0$ when $t$ is small.
By Weyl's inequality,
\begin{align}
0\le\lambda_j(L_t^\dagger N_0L_t+|t|^2F)
-\lambda_j(L_t^\dagger N_0L_t)
\le |t|^2\|F\|.
\end{align}
Together with the smoothness of the logarithm near $L_t^\dagger M_0L_t> 0$,
we have
\begin{align}
D(L_t^\dagger N_0L_t+|t|^2F\|
L_t^\dagger M_0L_t+|t|^2G)
-D(L_t^\dagger N_0L_t\|L_t^\dagger M_0L_t)
=O(|t|^2|\ln|t||).
\end{align}
Moreover,
$(1+|t|^2)^{-1}=1-|t|^2+O(|t|^4)$,
hence
\begin{align}
D(\N,\M;\tau_t)&=D(\N\|\M)
+2\Re\bigl(t\Tr\mathfrak B(N_0,M_0)Z\bigr)
+O(|t|^2|\ln|t||).
\end{align}
Optimality for every phase of $t$
gives $\Tr\mathfrak B(N_0,M_0)Z=0$.
\end{proof}

\begin{proof}[Proof of the inward bound \eqref{eq:qubit-inward-bound}]
Retain $\tau_p =\operatorname{diag}(1-p,p),\rho_p,\sigma_p$
from Section~\ref{app:alignment}, and recall
\begin{align}
L_p=\begin{pmatrix}\sqrt{1-p}I\\
\sqrt p\,Z^\dagger\end{pmatrix},\qquad
H_p=L_p^\dagger L_p=I+p(ZZ^\dagger-I),
\qquad
P_p=I-L_pH_p^{-1}L_p^\dagger.
\end{align}
By the alignment identities in \eqref{eq:qubit-block-alignment},
\begin{align}
\rho_p=L_pN_0L_p^\dagger+p(0\oplus F),
\qquad
\sigma_p=L_pM_0L_p^\dagger+p(0\oplus G).
\end{align}

Since $P_p$ and $I-P_p$ are complementary
orthogonal projections, pinching and
data processing give
\begin{align}
D(\N,\M;\tau_p)
&=D(\rho_p\|\sigma_p)\notag\\
&\ge D((I-P_p)\rho_p(I-P_p)\|
       (I-P_p)\sigma_p(I-P_p))
+D(P_p\rho_pP_p\|P_p\sigma_pP_p).
\end{align}

For the first pair, consider the isometry $V_p=L_pH_p^{-1/2}$ satisfying $V_pV_p^\dagger=I-P_p$. It identifies the first
compressed pair with
\begin{align}
V_p^\dagger\rho_pV_p
&=H_p^{1/2}N_0H_p^{1/2}
+p^2H_p^{-1/2}ZFZ^\dagger H_p^{-1/2},\notag\\
V_p^\dagger\sigma_pV_p
&=H_p^{1/2}M_0H_p^{1/2}
+p^2H_p^{-1/2}ZGZ^\dagger H_p^{-1/2}.
\end{align}
As in the preceding perturbation estimate,
the $O(p^2)$ corrections change the relative
entropy by $O(p^2|\ln p|)=o(p)$.
Hence the first pair contributes
$D_{N_0,M_0}(H_p)+o(p)$.

For the second pair, $P_pL_p=0$ gives
\begin{align}
P_p\rho_pP_p=pP_p(0\oplus F)P_p,\qquad
P_p\sigma_pP_p=pP_p(0\oplus G)P_p.
\end{align}
Moreover,
\begin{align}
(P_p)_{22}=I-pZ^\dagger H_p^{-1}Z=I+O(p).
\end{align}
Therefore, the isometry
\begin{align}
W_p=P_p\begin{pmatrix}0\\I\end{pmatrix}
(P_p)_{22}^{-1/2}
\end{align}
identifies the second compressed pair with
\begin{align}
W_p^\dagger\rho_pW_p
=p(P_p)_{22}^{1/2}F(P_p)_{22}^{1/2},
\qquad
W_p^\dagger\sigma_pW_p
=p(P_p)_{22}^{1/2}G(P_p)_{22}^{1/2}.
\end{align}

Since $(P_p)_{22}^{1/2}\to I$ and $F\le\lambda G$,
homogeneity and continuity of relative entropy
give $pD(F\|G)+o(p)$.

Combining both pairs yields
\begin{align}
D(\N,\M;\tau_p)\ge D_{N_0,M_0}(H_p)+pD(F\|G)+o(p).
\end{align}
Since $H_p=I+p(ZZ^\dagger-I)$,
the derivative in Lemma~\ref{lem:supporting-bound} and
$\Tr\mathfrak B(N_0,M_0)=D(\N\|\M)$ from
\eqref{eq:bregman-trace} give
\begin{align}
D_{N_0,M_0}(H_p)
&=D(\N\|\M)+p\Tr(ZZ^\dagger-I)\mathfrak B(N_0,M_0)+o(p)\notag\\
&=D(\N\|\M)+p\bigl[
\Tr Z^\dagger\mathfrak B(N_0,M_0)Z-D(\N\|\M)
\bigr]+o(p).
\end{align}
Consequently,
\begin{align}
D(\N,\M;\tau_p)&\ge D(\N\|\M)
+p\bigl[
\Tr Z^\dagger\mathfrak B(N_0,M_0)Z
+D(F\|G)-D(\N\|\M)
\bigr]+o(p).
\end{align}
Since $D(\N,\M;\tau_p)\le D(\N\|\M)$, the coefficient of $p$
is nonpositive, proving \eqref{eq:qubit-inward-bound}.
\end{proof}

\subsection{General block alignment}
\label{app:general-block-alignment}

\begin{boxlemma}[General block alignment]
\label{lem:general-block-alignment}
Let $A,B$ be finite-dimensional systems, and let
$\N,\M$ be channels from $A$ to $B$ with
$J_{\N},J_{\M}>0$. Suppose a state $\tau$
maximizes the one-copy objective in
\eqref{eq:input-objective}.
Let $P$ be the orthogonal projection onto $\supp\tau$.
Assume $P\ne I$ and set $Q=I-P$.
Relative to $(\ran P\oplus\ran Q)\otimes B$, write
\begin{align}
J_{\N}=\begin{pmatrix}N_0&X\\X^\dagger&N_1\end{pmatrix},
\qquad
J_{\M}=\begin{pmatrix}M_0&Y\\Y^\dagger&M_1\end{pmatrix}.
\end{align}
Then, with $Z:=M_0^{-1}Y$, we have $X=N_0Z$.
\end{boxlemma}

\begin{proof}
Choose a positive-definite state $\eta$ on $\ran Q$ and put
$\tau_u=(1-u)\tau+u\eta$. The corresponding outputs are
\begin{align}
\rho_u=\begin{pmatrix}
(1-u)c&\sqrt{u(1-u)}x\\
\sqrt{u(1-u)}x^\dagger&ur
\end{pmatrix},\qquad
\sigma_u=\begin{pmatrix}
(1-u)e&\sqrt{u(1-u)}y\\
\sqrt{u(1-u)}y^\dagger&us
\end{pmatrix},
\end{align}
where
\begin{align}
c&=(\sqrt\tau\otimes I_{d_B})N_0(\sqrt\tau\otimes I_{d_B}),
&e&=(\sqrt\tau\otimes I_{d_B})M_0(\sqrt\tau\otimes I_{d_B}),\notag\\
x&=(\sqrt\tau\otimes I_{d_B})X(\sqrt\eta\otimes I_{d_B}),
&y&=(\sqrt\tau\otimes I_{d_B})Y(\sqrt\eta\otimes I_{d_B}),\notag\\
r&=(\sqrt\eta\otimes I_{d_B})N_1(\sqrt\eta\otimes I_{d_B}),
&s&=(\sqrt\eta\otimes I_{d_B})M_1(\sqrt\eta\otimes I_{d_B}).
\end{align}
All inverses below are taken on $\ran P\otimes\mathbb C^{d_B}$.
Positive definiteness of the Choi matrices gives $c,e>0$ on
$\ran P\otimes\mathbb C^{d_B}$. The Schur complements on $\ran Q\otimes\mathbb C^{d_B}$ satisfy
\begin{align}
F=r-x^\dagger c^{-1}x>0,\qquad
G=s-y^\dagger e^{-1}y>0.
\end{align}
As in Appendix~\ref{app:alignment}, set
\begin{align}
R_u=\begin{pmatrix}\sqrt{1-u}\,I\\
\sqrt u\,x^\dagger c^{-1}\end{pmatrix},\qquad
L_u=\begin{pmatrix}\sqrt{1-u}\,I\\
\sqrt u\,y^\dagger e^{-1}\end{pmatrix}.
\end{align}
Then
\begin{align}
\rho_u=R_ucR_u^\dagger+u(0\oplus F),\qquad
\sigma_u=L_ueL_u^\dagger+u(0\oplus G).
\end{align}
Let $\Pi_u=I-L_u(L_u^\dagger L_u)^{-1}L_u^\dagger$.
Since $\Pi_uL_u=0$ and $(\Pi_u)_{22}=I+O(u)$,
\begin{align}
\Tr\rho_u\Pi_u
=u\Tr\!\left[F+(c^{-1}x-e^{-1}y)^\dagger
c(c^{-1}x-e^{-1}y)\right]+O(u^2).
\end{align}
Indeed, $R_u-L_u$ has only a lower block, equal
to $\sqrt u\,(c^{-1}x-e^{-1}y)^\dagger$.

Since $R_u^\dagger R_u=I+O(u)$, the nonzero eigenvalues
of $R_ucR_u^\dagger$ differ from those of $c$ by $O(u)$.
The trace-log estimate from Appendix~\ref{app:alignment} gives
\begin{align}
\Tr\rho_u\ln\rho_u
\ge\Tr c\ln c+u\ln u\,\Tr F-O(u).
\end{align}
For the other term, choose $\lambda_0>\|G\|$ and note that
$\sigma_u\le L_ueL_u^\dagger+\lambda_0uI$.
The latter equals $\lambda_0uI$ on $\ran\Pi_u$,
while its compression to $\ran L_u$ is $e+O(u)>0$.
Operator monotonicity of $\ln$ therefore gives
\begin{align}
\Tr\rho_u\ln\sigma_u
&\le\Tr c\ln e+(\ln u)\Tr\rho_u\Pi_u+O(u)\notag\\
&=\Tr c\ln e+u\ln u\,\Tr\!\left[
F+(c^{-1}x-e^{-1}y)^\dagger
c(c^{-1}x-e^{-1}y)\right]+O(u).
\end{align}
Subtracting, and using the trace-corrected definition
of relative entropy, we obtain
\begin{align}
&D(\N,\M;\tau_u)-D(\N,\M;\tau)
\ge u\ln(1/u)\Tr(x-ce^{-1}y)^\dagger
c^{-1}(x-ce^{-1}y)-O(u).
\end{align}
Optimality forces the nonnegative coefficient
to vanish, so $x=ce^{-1}y$. Since $\tau$ and $\eta$
are positive definite on their supports,
$X=N_0M_0^{-1}Y=N_0Z$.
\end{proof}

\end{document}